\documentclass[aps,prd,preprintnumbers,titlepage,nofootinbib,preprint]{revtex4-2}
\pdfoutput=1

\usepackage{amsmath,amsfonts}
\usepackage{hyperref}
\usepackage{tensor}
\usepackage{xcolor}

\newcommand{\fft}[2]{\frac{#1}{#2}}
\newcommand{\ft}[2]{{\textstyle\frac{#1}{#2}}}
\newcommand{\nn}{\nonumber}

\DeclareMathOperator{\Tr}{Tr}

\begin{document}
\preprint{LCTP-21-21}

\title{T-duality and hints of generalized geometry in string \texorpdfstring{$\alpha'$}{alpha'} corrections}

\author{Marina David}
\email{marina.david@kuleuven.be}
\affiliation{Instituut voor Theoretische Fysica, KU Leuven Celestijnenlaan 200D, B-3001 Leuven, Belgium}

\author{James T. Liu}
\email{jimliu@umich.edu}
\affiliation{Leinweber Center for Theoretical Physics, University of Michigan, Ann Arbor, MI 48109, U.S.A.}

\date{\today}

\begin{abstract}
We examine the structure of higher-derivative string corrections under a cosmological reduction and make connection to generalized geometry and T-duality. We observe that the natural T-duality quantities are a linear combination of the $B$-field and graviton, which readily appears in the torsionful connection, $\Omega_\pm=\Omega\pm\fft12H$.  We revisit the tree-level $\alpha'R^2$ corrections to the bosonic and heterotic string using the $\Omega_\pm$ framework. We then turn to the structure of the T-duality completion of tree-level $\alpha'^3R^4$ in the type II string and validate results in literature obtained for the tree-level five-point contact terms of the form $H^2R^3$.
\end{abstract}

\maketitle

\section{Introduction}

As a natural candidate for a UV complete theory of quantum gravity, string theory allows us to go beyond its low-energy supergravity limit.  Perturbative stringy corrections to supergravity can be organized through both the $\alpha'$ expansion and the string loop (genus) expansion.  From an effective field theory point of view, the $\alpha'$ expansion corresponds to higher derivative corrections, and such corrections can show up at any order in the genus expansion.

In a covariant theory of gravity, the higher derivative corrections take the form of an expansion in curvature, with couplings of the form $(\alpha')^{m+n-1}D^{2m}R^n$.  For the bosonic and heterotic string, this expansion starts at order $\alpha'R^2$, while maximal supersymmetry of the type II string requires the expansion to start at order $\alpha'^3R^4$.  Such couplings can be probed starting from four-point string amplitudes and have been extensively studied since the pioneering work of \cite{Schwarz:1982jn,Gross:1986iv,Sakai:1986bi,Gross:1986mw}.  For the type II string, amplitudes involving NSNS sector fields can be specified by an arbitrary closed string polarization tensor $\theta_{\mu\nu}$ with symmetric trace-free, antisymmetric, and trace components corresponding to the individual fields $g_{\mu\nu}$, $B_{\mu\nu}$, and $\phi$, respectively.  As a result, the four-point function results can be compactly written using the linearized Riemann tensor with torsion \cite{Gross:1986iv,Gross:1986mw}
\begin{equation}
    \bar R_{\mu\nu}{}^{\rho\sigma}(\Omega_\pm)=R_{\mu\nu}{}^{\rho\sigma}(\Omega)\pm\nabla_{[\mu}H_{\nu]}{}^{\rho\sigma},
\label{eq:R+lin}
\end{equation}
where
\begin{equation}
    \Omega_{\pm\,\mu}{}^{\rho\sigma}=\Omega_\mu{}^{\rho\sigma}\pm\fft12H_\mu{}^{\rho\sigma}.
\label{eq:Omega+}
\end{equation}
The type II string receives perturbative eight-derivative corrections at both tree-level and one-loop, and the quartic effective action takes the form
\begin{widetext}
\begin{align}
    e^{-1}\mathcal L&=e^{-2\phi}\left(R+4\partial_\mu\phi^2-\fft1{12}H_{\mu\nu\rho}^2+\frac{\zeta(3)}{3\cdot2^{11}}(t_8t_8-\ft14\epsilon_8\epsilon_8)\bar R(\Omega_+)^4+\cdots\right)\nn\\
    &\quad+\left(\fft{\pi^2}{9\cdot2^{11}}(t_8t_8\pm\ft14\epsilon_8\epsilon_8)\bar R(\Omega_+)^4+\cdots\right),
\label{eq:quartA}
\end{align}
\end{widetext}
in the NSNS sector.  In the second line, the top (bottom) sign corresponds to type IIA (IIB) string theory.

From a stringy perspective, the connections $\Omega_+$ and $\Omega_-$ are related by world-sheet parity, and neither is preferred.  In particular, in the type II string with $dH=0$, the torsionful Riemann tensor satisfies $R_{\mu\nu\rho\sigma}(\Omega_+)=R_{\rho\sigma\mu\nu}(\Omega_-)$ even at the full non-linear level.  This allows us to use a convention where torsionful expressions are written using only the $\Omega_+$ connection, as for example we have done in (\ref{eq:quartA}).  However, it should be understood, at least for type II strings, that $\Omega_+$ and $\Omega_-$ enter on an equal footing%
\footnote{The heterotic string case is a bit different, as world-sheet parity is no longer applicable.  Here there is an intricate interplay between $\Omega_+$ an $\Omega_-$ in the bosonic effective action and the supersymmetry variations \cite{Bergshoeff:1988nn,Bergshoeff:1989de} as required by supersymmetry.}.

The use of the connection with torsion, (\ref{eq:Omega+}), was emphasized in the heterotic framework in \cite{Bergshoeff:1989de}, and furthermore the parametrization of NSNS amplitudes in terms of $\theta_{\mu\nu}$ is suggestive of double field theory \cite{Hull:2009mi,Hohm:2010pp} and generalized geometry \cite{Coimbra:2011nw}.  This connection is perhaps even stronger when the action is compactified on a $d$-dimensional torus, as the T-duality group $O(d,d)$ is naturally identified as the structure group of the generalized tangent bundle.  Hence T-duality acts naturally on the generalized metric on $T^d$.

There is a remarkable amount of structure in the quartic effective action (\ref{eq:quartA}).  However, it is far from complete.  In particular, four-point string scattering only probes the linearized Riemann tensor with torsion, while general eight-derivative contact terms will extend to the level of eight-point functions, such as $\partial\phi^8$ and $H^8$.  Furthermore, at the four-point level, there will also be couplings involving RR fields \cite{Policastro:2006vt,Policastro:2008hg}.  Even without a detailed computation, the RR contributions must be present because of supersymmetry, as the entire massless sector of type II theory resides in a single multiplet.

While the full structure of the eight derivative couplings in type II theory in the form of a completion of (\ref{eq:quartA}) is still unknown, progress has been made along multiple directions.  Perhaps the most straightforward way to proceed is to compute higher-point string amplitudes.  Note, however, that while such amplitudes are generally under good control at the tree and one-loop level, recreating an effective action from them requires additional work.  The one-loop five-point effective action involving gravitons and the $H$-field was investigated in \cite{Peeters:2001ub,Richards:2008jg,Richards:2008sa,Liu:2013dna}, and the tree-level $H^2R^3$ couplings were obtained in \cite{Liu:2019ses}.  In principle, it would be a straightforward exercise to complete the quintic effective action by including dilaton couplings.  However, working to six-points and higher will remain a challenge as extracting the contact interactions from the amplitude will require numerous subtractions.

Because of the difficulties of the direct approach to recreating the eight-derivative effective action, complementary ways have been developed at arriving at the couplings.  One natural idea is to make use of supersymmetry in the sense that there ought to be a natural supersymmetric completion of the $\alpha'^3R^4$ coupling that makes use of the full type II supergravity multiplet.  In this case, we would expect different supersymmetric completions depending on whether we consider type IIA or type IIB theory.  Nevertheless, at tree level, the NSNS sector is universal and can be organized into $\mathcal N=1$ superinvariants.  Formally, writing the higher curvature couplings in terms of superinvariants is perhaps the most elegant formulation.  However, working with ten-dimensional supersymmetry can be a challenge, and only partial results have been obtained along these lines (see \textit{e.g.}\ \cite{Green:1998by,Cederwall:2000ye,Peeters:2000qj,Gates:2001hf,Nishino:2001mb,Cederwall:2004cg,Rajaraman:2005ag,Paulos:2008tn,Becker:2017zwe,Becker:2021oiz}).

Another fruitful approach to elucidating the structure of the higher derivative action is to make use of stringy symmetries including T-duality and, for the type IIB string, S-duality.  In fact, S-duality is highly constraining as it groups together tree level, one-loop and non-perturbative $\alpha'^3R^4$ terms in order to form an $SL(2,\mathbb Z)$ invariant action \cite{Green:1997tv,Green:1997di,Green:2019rhz}.  When rewritten in the Einstein frame, the higher-derivative type IIB graviton couplings in (\ref{eq:quartA}) take the form
\begin{equation}
    e^{-1}\mathcal L_{\text{II}}^{\partial^8}\sim \alpha'^3f_0(\tau,\bar\tau)(t_8t_8-\ft14\epsilon_8\epsilon_8)R^4+\cdots,
\end{equation}
where $f_0(\tau,\bar\tau)$ is a non-holomorphic Eisenstein series of weight 3/2.  The full set of quartic couplings preserve the $U(1)$ R-symmetry of type IIB supergravity, while higher-point couplings can break it.  However, at each order in the number of fields, there is a maximum bound on the amount of $U(1)$ violation.  Maximal $U(1)$ violating couplings have a relatively simpler structure than those that do not saturate the bound \cite{Boels:2012zr,Green:2019rhz}, and can in principle be worked out without too much difficulty.  However, getting a handle on non-maximal $U(1)$ violating couplings still remains somewhat of a challenge.

In addition to S-duality, we can consider the constraints on the effective action imposed by T-duality invariance.  At a superficial level, T-duality interchanges `momentum' with `winding' modes, which corresponds, in an effective supergravity point of view, to the interchange of $g_{\mu9}\leftrightarrow b_{\mu9}$.  While T-duality maps between type IIA and type IIB theories, this distinction is unimportant when restricted to the NSNS sector.  Of course, T-duality investigations of the effective supergravity action are generally restricted to the tree-level couplings, as both Kaluza-Klein and string winding modes will be important at the loop level and will lead to additional complications when transforming the one-loop couplings.  Since tree-level NSNS couplings do not distinguish between type IIA and type IIB theory, T-duality invariance in this sector is universal in the type II string.

The $\mathcal O(\alpha')$ corrections to the T-duality transformation rules were investigated in \cite{Bergshoeff:1995cg,Kaloper:1997ux}, and a two-parameter family of T-duality invariant completions of $\alpha'R^2$ was constructed in \cite{Marques:2015vua,Baron:2017dvb}.  An appropriate choice of parameters then interpolates between corrections to the bosonic string and to the heterotic string.  T-duality invariance of $\mathcal O(\alpha')$ corrected black holes as well as the action have been investigated in \cite{Cano:2018qev,Edelstein:2018ewc,Cano:2018brq,Edelstein:2019wzg,Elgood:2020xwu,Ortin:2020xdm}.  T-duality of the $\mathcal O(\alpha'^3)$ corrections to type~II theory was explored in \cite{Garousi:2012yr,Garousi:2012jp,Garousi:2013zca,Liu:2013dna,Liu:2019ses}, and has been used to put constraints on additional couplings with the $H$-field.

While \cite{Marques:2015vua,Baron:2017dvb} presents a two-parameter family of T-duality invariant $\mathcal O(\alpha')$ actions, only specific choices of the parameters correspond to the heterotic and bosonic strings.  Thus T-duality invariance uniquely determines the completion of the $\alpha'R^2$ coupling in the respective string theories.  It was further shown in \cite{Garousi:2019wgz,Garousi:2019mca} that this extends to the uniqueness of $\alpha'^2R^3$ couplings in the bosonic string.  Recently this has been extended to the type~II context where once again T-duality invariance on a circle fixes the entire sector of eight-derivative NSNS couplings up to an overall coefficient \cite{Garousi:2020gio,Garousi:2020lof}.

In principle, the full NSNS completion of the effective action (\ref{eq:quartA}) is now known from T-duality.  However, the construction of \cite{Garousi:2020gio} makes use of the minimal basis developed in \cite{Garousi:2020mqn}, which is not readily comparable with the form of the effective action given in (\ref{eq:quartA}). Both forms of the action have been shown to be equivalent at the level of four-point NSNS interactions \cite{Garousi:2020gio,Garousi:2020lof}. One of our goals in this work to extend this matching of the level of the five-point $H^2R^3$ interactions given in \cite{Liu:2019ses}.  Since the eight-derivative action is fully constrained by T-duality, an appropriate field redefinition can be made to put the action of \cite{Garousi:2020gio,Garousi:2020lof} in the form of (\ref{eq:quartA}). But, in practice, this is not an easy task.  Moreover, the structure of the basis used in either \cite{Garousi:2020mqn,Garousi:2020gio} or \cite{Garousi:2020lof} is not particularly illuminating from a string geometry point of view.  Thus another goal is to better understanding T-duality and the higher derivative corrections from a generalized geometry point of view and to take a closer look the completion of $\alpha'^3R^4$ to see if it is possible to interpret the result of \cite{Garousi:2020gio,Garousi:2020lof} more directly from a world-sheet perspective.

While T-duality in the effective action is perhaps most closely associated with the interchange of `momentum' and `winding' gauge fields, $O(d,d)$ invariance also applies to the scalars $g_{ij}$ and $b_{ij}$ where $i$ and $j$ are $T^d$ indices.  This invariance was used successfully to test and constrain the higher derivative corrections to the bosonic string in \cite{Meissner:1996sa,Godazgar:2013bja}.  Subsequently, a systematic approach to T-duality invariance in the scalar sector was developed in \cite{Hohm:2015doa}.  The general idea is to perform a `cosmological' reduction by compactifying all spatial directions so that the only remaining fields are the time-dependent scalars $g_{ij}(t)$ and $b_{ij}(t)$.  T-duality invariance then imposes strong conditions on the possible couplings of the reduced scalars.

The cosmological reduction idea of \cite{Meissner:1996sa,Godazgar:2013bja,Hohm:2015doa} provides a systematic method of testing T-duality invariance of higher derivative couplings.  Recently, it was shown in \cite{Codina:2020kvj} that the tree-level eight-derivative graviton couplings in (\ref{eq:quartA}), which can be expressed as
\begin{align}
    e^{-1}\mathcal L_{R^4}&\sim (t_8t_8-\ft14\epsilon_8\epsilon_8)R^4\nn\\
    &=-192(R_{\mu\nu}{}^{\rho\sigma}R^{\mu\nu\lambda\eta}R_{\lambda\rho}{}^{\zeta\xi}R_{\eta\sigma\zeta\xi}\nn\\
    &\kern4em-4R^{\mu\nu\rho\sigma}R_\mu{}^\eta{}_\rho{}^\lambda R_\nu{}^\zeta{}_\eta{}^\xi R_{\sigma\zeta\lambda\xi})+\cdots,
\end{align}
reduces to
\begin{equation}
    e^{-1}\mathcal L_{R^4}\ \to\ -9\Tr(L^8)+6\left(\Tr(L^4)\right)^2,
\end{equation}
with $L^i{}_j=g^{ik}\dot g_{kj}$, where a dot denotes the time derivative $d/dt$, and provided the Ricci terms in ellipses are discarded.  As noted in \cite{Hohm:2015doa}, this expression is compatible with T-duality as it does not contain any traces of odd powers of $L$ that would explicitly break T-duality invariance.

While compatibility with T-duality invariance can be tested even without the antisymmetric tensor fields, a stronger test of full $O(9,9)$ invariance would necessarily involve both $g_{ij}$ and $b_{ij}$.  Along these lines, it was shown in \cite{Garousi:2021ikb} that the full set of eight-derivative couplings obtained from T-duality invariance on $S^1$ obtained in \cite{Garousi:2020gio,Garousi:2020lof} indeed reduces to the complete T-duality invariant expression
\begin{equation}
    e^{-1}\mathcal L_{\partial^8}\ \to\ -\fft92\Tr\relax(\dot{\mathcal S}^8)+\fft32\Tr\relax(\dot{\mathcal S}^4)^2,
\label{eq:ans}
\end{equation}
where $\mathcal S$ is defined in \cite{Hohm:2015doa} and below in (\ref{eq:Sdef}).

Since the eight-derivative effective action in \cite{Garousi:2020gio,Garousi:2020lof} was obtained by demanding invariance under T-duality, its cosmological reduction to (\ref{eq:ans}) would necessarily have to be invariant by construction.  However, it remains instructive to see how the different higher derivative terms assemble themselves to form supersymmetric and T-duality invariant combinations.  Along the same lines, we may hope for a deeper understanding of the structure of stringy higher derivative corrections and what role generalized geometry may play in formulating T-duality invariant couplings.  With this in mind, we revisit the cosmological reduction of \cite{Hohm:2015doa} making direct use of the connection with torsion (\ref{eq:Omega+}).

We show that the cosmological reduction of \cite{Hohm:2015doa} has an elegant formulation when written in terms of the suggestive combination $N^i{}_{j\pm}=g^{ik}(\dot g_{kj}\pm\dot b_{kj})$.  By analyzing the reduction of the torsionful Riemann tensor $R_{\mu\nu}{}^{\rho\sigma}(\Omega_\pm)$, we see hints of generalized geometry in the higher derivative couplings. At the same time, we demonstrate that complete tree-level invariants will involve the $H$ field and dilaton.

This paper is organized as follows. In the next section we review the cosmological reduction of \cite{Hohm:2015doa} and present a reformulation in terms of the metric and $H$-field combination $N^i{}_{j\pm}$.  We demonstrate that T-duality invariant expressions take the form of traces of an even number of $N$'s alternating between $N_+$ and $N_-$.  In section~\ref{sec:hciR} we reduce the torsionful Riemann tensor $R_{\mu\nu}{}^{\rho\sigma}(\Omega_\pm)$ and show that with additional couplings to the $H$-field, the T-duality invariant quantities arise as linear combinations of the B-field and graviton.  We then turn to eight-derivative couplings in the Type II string in section~\ref{sec:apm3} and provide a check on the five-point $H^2R^3$ contact terms obtained in \cite{Liu:2019ses}.  Finally, we make some concluding remarks in section~\ref{sec:conc}.  In the Appendix, we give the basis used for the $O(9,9)$ invariant completion of the eight-derivative couplings up to order $H^2R^3$.

\section{Cosmological reduction and T-duality}
\label{sec:cred}

Before turning to a reexamination of the eight-derivative couplings in the type II effective action, we review the cosmological reduction of \cite{Meissner:1996sa,Godazgar:2013bja,Hohm:2015doa} and the investigation of T-duality invariance in the scalar sector of the reduced theory.  Here the focus is on the tree-level effective action, so that we can avoid introducing Kaluza-Klein and winding modes on the torus.  We further restrict to the closed string NSNS fields, in which case we need not make a distinction between type IIA and type IIB theory.

The tree-level NSNS couplings are universal and involve the massless fields $(G_{\mu\nu},B_{\mu\nu},\phi)$ with leading two-derivative low energy effective Lagrangian
\begin{equation}
    e^{-1}\mathcal L_{10}=e^{-2\phi}\left(R+4\partial_\mu\phi\partial^\mu\phi-\fft1{12}H_{\mu\nu\rho}H^{\mu\nu\rho}\right).
\end{equation}
Reduction on $T^d$ proceeds by making the ansatz
\begin{align}
    ds_{10}^2&=g_{\alpha\beta}dx^\alpha dx^\beta+g_{ij}(dy^i+A_\alpha^i dx^\alpha)(dy^j+A_\beta^jdx^\beta),\nn\\
    B&=\ft12\hat B_{\alpha\beta}dx^\alpha\wedge dx^\beta+B_{\alpha i}dx^\alpha\wedge(dy^i+A_\beta^i dx^\beta)\nn\\
    &\qquad+\ft12b_{ij}(dy^i+A_\alpha^i dx^\alpha)\wedge(dy^j+A_\beta^jdx^\beta),\nn\\
    \phi&=\ft12\Phi+\ft14\log\det g_{ij},
\label{eq:KKans}
\end{align}
where $\alpha,\beta,\gamma$ denote lower-dimensional indices on the $(10-d)$ space and $i,j$ denote indices on the $T^{d}$.
The resulting torus-reduced two-derivative Lagrangian takes the from
\begin{widetext}
\begin{equation}
    e^{-1}\mathcal L_{10-d}=e^{-\Phi}\left(R+\partial_\alpha\Phi\partial^\alpha\Phi-\fft1{12}\hat H_{\alpha\beta\gamma}\hat H^{\alpha\beta\gamma}+\fft18\Tr(\partial_\alpha\mathcal H\eta\partial^\alpha\mathcal H\eta)-\fft14\mathcal F_{\alpha\beta}\eta\mathcal H\eta\mathcal F^{\alpha\beta}\right).
\label{eq:L10-d}
\end{equation}
\end{widetext}
Here $\mathcal H$ is the $(2d)\times(2d)$ scalar matrix
\begin{equation}
    \mathcal H=\begin{pmatrix}g^{-1}&-g^{-1}b\\bg^{-1}&g-bg^{-1}b\end{pmatrix},
\label{eq:ggmet}
\end{equation}
satisfying $\mathcal H^{-1}=\eta\mathcal H\eta$ where $\eta$ is the $O(d,d)$ metric
\begin{equation}
    \eta=\begin{pmatrix}0&\mathbf1\\\mathbf1&0\end{pmatrix}.
\end{equation}
Note that $g$ and $b$ denote the scalars $g_{ij}$ and $b_{ij}$, respectively.  From a geometrical point of view, $\mathcal H^{-1}$ is the generalized metric on $T^d$.  In addition, the `momentum' and `winding' gauge fields have been grouped together according to
\begin{equation}
    \mathcal A_\alpha=\begin{pmatrix}A_\alpha^i\\B_{\alpha i}\end{pmatrix},\qquad\mathcal F_{\alpha\beta}=\partial_{[\alpha}\mathcal A_{\beta]}.
\end{equation}

The reduced two-derivative Lagrangian (\ref{eq:L10-d}) is invariant under $O(d,d)$ T-duality transformations of the form
\begin{equation}
    \mathcal H\to\Omega\mathcal H\Omega^T,\qquad\mathcal A_\alpha\to\Omega\mathcal A_\alpha,
\label{eq:TMA}
\end{equation}
where $\Omega^T\eta\Omega=\eta$.  While we only consider the NSNS sector, T-duality continues to be a symmetry with the addition of the RR sector and is in fact enhanced to a larger U-duality symmetry \cite{Cremmer:1978ds,Cremmer:1979up,Hull:1994ys}.  Although the full U-duality is non-perturbative, T-duality is perturbative and will persist at higher orders in the derivative expansion.

Despite the fact that the full power of T-duality involves the combined transformation of scalars and vectors as indicated in (\ref{eq:TMA}), a surprising amount of information can already be obtained from the scalar sector alone, as demonstrated in \cite{Meissner:1996sa,Godazgar:2013bja,Hohm:2015doa,Hohm:2019jgu,Codina:2020kvj,Garousi:2021ikb,Codina:2021cxh}.  Focusing only on the scalar sector, T-duality invariant couplings can be constructed out of the lower-dimensional dilaton $\Phi$, the scalar matrix $\mathcal H$, and their derivatives.  While $\Phi$ is invariant under $O(d,d)$ transformations, the $O(d,d)$ scalars transform according to
\begin{equation}
    (\mathcal H\eta)\to\Omega(\mathcal H\eta)\Omega^{-1},
\end{equation}
as can be deduced from (\ref{eq:TMA}).  T-duality invariant combinations then take the form of traces
\begin{equation}
    \Tr(\partial^{n_1}(\mathcal H\eta)\partial^{n_2}(\mathcal H\eta)\cdots),
\end{equation}
where $\partial^n$ schematically denotes any number of derivatives with appropriate contractions of their Lorentz indices.  At the two-derivative level, this is clearly seen in the scalar kinetic term in (\ref{eq:L10-d}).

While the scalar sector can be investigated in any reduced dimension, perhaps the cleanest approach is that of \cite{Meissner:1996sa,Hohm:2015doa}, which is to perform a `cosmological' reduction on $T^9$ to arrive at a one-dimensional theory with only time-dependent scalars $(g_{ij}, b_{ij},\Phi)$. In this case, the reduction ansatz (\ref{eq:KKans}) takes the form
\begin{align}
\label{graviton reduction}
    G_{\mu\nu} &= \begin{pmatrix} -n(t)^2 & 0 \\ 0 & g_{ij}(t) \end{pmatrix},\qquad B_{\mu\nu}=\begin{pmatrix} 0 && 0 \\ 0 && b_{ij}(t) \end{pmatrix},\nn\\
    \qquad\phi &= \ft{1}{2}\Phi(t) + \ft{1}{4}\log \det g_{ij}(t),
\end{align}
and all gauge fields including the antisymmetric tensor $\hat B_{\alpha\beta}$ are absent in the compactified theory.  The reduced gravitational sector is also trivial, with only the lapse function $n(t)$ remaining, and even that can be removed by time reparametrization.  At the two-derivative level, the one-dimensional Lagrangian (\ref{eq:L10-d}) takes the simple form
\begin{equation}
    \mathcal L_1=\fft{e^{-\Phi}}n\left(-\dot\Phi^2-\ft18\Tr\relax(\dot{\mathcal S}^2)\right),
\label{eq:lag1d}
\end{equation}
where dots denote time derivatives $d/dt$ and we have defined
\begin{equation}
    \mathcal S=\eta\mathcal H,
\label{eq:Sdef}
\end{equation}
following \cite{Hohm:2015doa}.  Note that the lapse function $n(t)$ in the denominator in \eqref{eq:lag1d} ensures the time-reparametrization invariance of the action, $S=\int\mathcal L_1dt$.

Higher-derivative invariants in the reduced theory can be constructed out of powers of $\Phi$ and traces of $\mathcal S^n$ along with their covariant time derivatives.  Using reparametrization invariance, we can set $n(t)=1$, which we will do from now on.  Furthermore, as demonstrated in \cite{Hohm:2015doa}, use of field redefinitions and on-shell equations of motion allow all higher-derivative $O(d,d)$ invariants to be written in the form of traces of even powers of $\dot{\mathcal S}$
\begin{equation}
    \Tr\relax(\dot{\mathcal S}^2),\qquad\Tr\relax(\dot{\mathcal S}^4),\qquad\Tr\relax(\dot{\mathcal S}^6),\qquad\ldots.
\label{eq:TrSdn}
\end{equation}
At the four-derivative level, the T-duality invariant couplings can thus be written as
\begin{equation}
    \mathcal L_{\partial^4}=A_1\Tr\relax(\dot{\mathcal S}^2)^2+A_2\Tr\relax(\dot{\mathcal S}^4),
\label{eq:L4Tdi}
\end{equation}
while at the eight-derivative level, they can be parametrized as
\begin{align}
    \mathcal L_{\partial^8}&=A_1\Tr\relax(\dot{\mathcal S}^4)^2+A_2\Tr\relax(\dot{\mathcal S}^8)+A_3\Tr\relax(\dot{\mathcal S}^2)^4\nn\\
    &\quad+A_4\Tr\relax(\dot{\mathcal S}^2)^2\Tr\relax(\dot{\mathcal S}^4)+A_5\Tr\relax(\dot{\mathcal S}^2)\Tr\relax(\dot{\mathcal S}^6).
\label{8 derivative general action}
\end{align}
As shown in \cite{Hohm:2019jgu}, this basis of $O(d,d)$ invariants is non-minimal as powers of $\Tr\relax(\dot{\mathcal S}^2)$ can be eliminated by a further field redefinition.  Nevertheless, they are allowed as far as invariants are concerned.

Since the scalar matrix $\mathcal S$ is built out of $g_{ij}$ and $b_{ij}$, its time derivative $\dot{\mathcal S}$ will involve derivatives of these fields.  Following \cite{Hohm:2015doa}, we introduce the matrices
\begin{equation}
    L=g^{-1}\dot g,\qquad M=g^{-1}\dot b,
\end{equation}
which in components reads $L^i{}_j=g^{ik}\dot g_{kj}$ and $M^i{}_j=g^{ik}\dot b_{kj}$.  At the same time, however, from a stringy point of view, the scalars $g_{ij}$ and $b_{ij}$ naturally enter in the combination $\Theta_{ij}=g_{ij}+b_{ij}$.  This feature of the closed string NSNS sector leads us to define the linear combinations
\begin{equation}
    N_\pm=L\pm M\quad\mbox{or, in components}\quad N_\pm^i{}_j=g^{ik}(\dot g_{kj}\pm\dot b_{kj}).
\label{eq:Npm}
\end{equation}
Note that the matrices $L$, $M$, and $N_\pm$ are implicitly written with first index raised and second index lowered, so covariant expressions can be obtained by ordinary matrix multiplication without use of the metric $g_{ij}$ or its inverse.  Since $g_{ij}$ is symmetric while $b_{ij}$ is antisymmetric, $N_\pm$ satisfies the transpose relation
\begin{equation}
    N_\pm^T=gN_\mp g^{-1}.
\label{eq:Ntransp}
\end{equation}
One advantage of introducing $N_\pm$ is that the T-duality invariant traces (\ref{eq:TrSdn}) can be compactly written as
\begin{align}
    \Tr\relax(\dot{\mathcal{S}}^{2n}) = 2(-1)^{n}\Tr((N_{+}N_{-})^{n\vphantom1}),\qquad (n\ge0).
\label{eq:Tdi}
\end{align}

It follows from (\ref{eq:Tdi}) that the cosmological reduction allows for a non-trivial test of T-duality invariance.  The procedure begins by reducing the higher-derivative couplings to scalar interactions according to (\ref{graviton reduction}).  We can then express the result in terms of the matrices $N_\pm$ according to (\ref{eq:Npm}).  Generally covariant expressions will always reduce to products of traces, so verifying T-duality invariance becomes a matter of seeing whether these traces are all of the form (\ref{eq:Tdi}), where the string of matrices alternate between $N_+$ and $N_-$.

Note that there is a non-trivial step hidden in this procedure, as a straightforward reduction will generically lead to interaction terms involving time derivatives of the dilaton as well as higher time derivatives of $N_\pm$.  To facilitate comparison with (\ref{eq:Tdi}), it is useful to make use of on-shell field redefinitions and integration by parts in the reduced action to convert such terms into a canonical form that can be expressed entirely in terms of traces of $N_{\pm}$ with no further time derivatives and with decoupled dilaton.  We now describe how this can be done in general.

\subsection{Equations of motion and field redefinitions}
\label{sec:eomfr}

When investigating higher-derivative couplings, it is important to note that there is a lot of freedom in performing field redefinitions.  Thus it is often far from obvious whether two expressions are physically equivalent or not.  One way to manage this freedom is to construct a `minimal' basis up to field redefinitions.  However, while in some cases there may be a preferred basis, usually such a choice is somewhat arbitrary.

Focusing on the cosmological reduction on $T^9$, gauge invariant higher derivative invariants can be constructed out of $\Phi$ and $N_\pm$ and their time derivatives.  However, as shown in \cite{Hohm:2015doa,Codina:2021cxh}, the use of field redefinitions and on-shell equations of motion can eliminate all time derivatives of $\Phi$ and $N_\pm$, leading to a minimal basis consisting of traces of strings of $N_+$ and $N_-$.  In this case, the derivative counting is straightforward, as each $N_\pm$ counts precisely one time derivative.  Note that this basis is more general than that constructed out of the T-duality invariants (\ref{eq:TrSdn}). The cosmological reduction of any covariant higher-dimensional action can be brought into the $N_\pm$ basis.  Only if the original action is T-duality invariant can the resulting traces of $N_\pm$ be assembled into traces of $\dot{\mathcal S}$ according to (\ref{eq:Tdi}).

To highlight the field redefinitions that can be used to bring the higher-derivative action into a canonical form, we start with the two-derivative equations of motion.  A straightforward variation of the one-dimensional Lagrangian, (\ref{eq:lag1d}), gives rise to the set of equations (after setting $n(t)=1$)
\begin{align}
    \dot\Phi^2&=\ft14\Tr\relax(N_+N_-),\nn\\
    \ddot\Phi-\ft12\dot\Phi^2&=\ft18\Tr\relax(N_+N_-),\nn\\
    \dot N_\pm-\dot\Phi N_\pm&=\pm MN_\pm.
\label{EoM 1D}
\end{align}
Note that the first equation is obtained from varying (\ref{eq:lag1d}) with respect to the lapse function $n(t)$ before restricting to the $n(t)=1$ gauge.  It is easily verified that these equations are consistent with the reduction of the original ten-dimensional NSNS equations
\begin{align}
\label{EoM at O(1)}
    R-4 \partial \phi^{2}+4 \square \phi-\frac{1}{12} H_{\mu \nu \rho}^{2}&=0,\nn \\
    R_{\mu \nu}+2 \nabla_{\mu} \nabla_{\nu} \phi-\frac{1}{4} H_{\mu \rho \sigma} \tensor{H}{_{\nu}^{\rho \sigma}} &=0, \nn\\
    d\left(e^{-2 \phi} * H\right) &=0.
\end{align}

We now outline the procedure that can be used to write the higher derivative terms in a canonical form using the equations of motion and integration by parts%
\footnote{A similar systematic procedure is given in \cite{Codina:2021cxh} in the absence of the $H$-field.}.
The first step is to remove time derivatives of $\dot\Phi$ and $N_\pm$.  Removing time derivatives of $N_\pm$ is straightforwardly done using the final equation of (\ref{EoM 1D})
\begin{equation}
    \dot N_\pm\to(\dot\Phi\pm M)N_\pm,
\label{eq:Ndotelim}
\end{equation}
along with additional time derivatives of this expression, if necessary.  After eliminating all time derivatives of $N_\pm$, we then remove second and higher time derivatives of $\Phi$ by applying a combination of the first two equations of (\ref{EoM 1D})
\begin{equation}
    \ddot\Phi\to\dot\Phi^2.
\label{eq:Pdotelim}
\end{equation}
Note that both (\ref{eq:Ndotelim}) and (\ref{eq:Pdotelim}) increase the non-linear order of the fields in that they effectively replace one time derivative $d/dt$ by multiplication with either $\dot\Phi\pm M$ or $\dot\Phi$.

After eliminating additional time derivatives, we would be left with an expression given entirely in powers of $\dot\Phi$ multiplied by traces of $N_\pm$.  At this stage, there are various ways to proceed.  One way is to eliminate even powers of $\dot\Phi$ using the first equation in (\ref{EoM 1D}) until only a single power of $\dot\Phi$ remains.  The single power of $\dot\Phi$ can then be removed using integration by parts on the equations of motion \cite{Hohm:2015doa}.  However, this procedure by itself does not remove all field redefinition ambiguities, as terms involving $\Tr\relax(N_+N_-) = \Tr\relax(L^2-M^2)$ can still be shifted around using the first two equations of (\ref{EoM 1D}).

Note that two independent traces are possible at the quadratic level
\begin{align}
    \Tr\relax(N_+N_-)&=\Tr\relax(L^2-M^2),\nn\\
    \Tr\relax(N_+^2)&=\Tr\relax(N_-^2)=\Tr\relax(L^2+M^2).
\end{align}
Following \cite{Codina:2020kvj,Codina:2021cxh}, it is convenient to choose a canonical basis that eliminates all powers of $\Tr\relax(L^2)$ so that the basis becomes minimal when the $H$-field is truncated out (\textit{i.e.}\ when $M\to0$).  Along these lines, we make the substitution%
\footnote{Alternatively, a more streamlined procedure would be to make the first substitution for $\Tr(N_+N_-)$ in (\ref{eq:N2subs}), but to omit the other two.  However, this will leave $\Tr(L^2+M^2)$ in the canonical basis.  The choice made here is needed to remove \emph{all} $\Tr(L^2)$ terms from the basis.}
\begin{align}
    \Tr\relax(N_+N_-)&\to4\dot\Phi^2,\nn\\
    \Tr(N_+^2)&\to4\dot\Phi^2+2\Tr\relax(M^2),\nn\\
    \Tr(N_-^2)&\to4\dot\Phi^2+2\Tr\relax(M^2).
\label{eq:N2subs}
\end{align}
At this point, all $n$-derivative terms have been reduced to expressions of the form
\begin{equation}
    \dot\Phi^k F_{n-k}(N_+,N_-),
\end{equation}
where $F_{n-k}(N_+,N_-)$ is homogeneous of degree $n-k$ and does not include any powers of $\Tr(L^2)$.  Note that $F_{n-k}(N_+,N_-)$ can include the linear trace $\Tr N_+=\Tr N_-=\Tr L$ and the quadratic trace $\Tr\relax(M^2)$, which can be expressed as the combination $\fft12(\Tr(N_+^2)-\Tr(N_+N_-))$, along with any combination of cubic and higher traces.

To remove powers of $\dot\Phi$, we make use of integration by parts, which amounts to adding a total derivative to the Lagrangian of the form
\begin{widetext}
\begin{align}
    \fft{d}{dt}\left(e^{-\Phi}\dot\Phi^{k-1}F_{n-k}(N_+,N_-)\right)&\nn\\
    &\kern-8em=e^{-\Phi}\left(\left(-\dot\Phi^k+(k-1)\dot\Phi^{k-2}\ddot\Phi\right)F_{n-k}(N_+,N_-)+\dot\Phi^{k-1}\fft{d}{dt}F_{n-k}(N_+,N_-)\right).
\label{eq:totder}
\end{align}
The $\ddot\Phi$ term can be replaced using the equation of motion substitution (\ref{eq:Pdotelim}), while the total time derivative of $F_{n-k}(N_+,N_-)$ can be evaluated using the chain rule along with the substitution (\ref{eq:Ndotelim}).  Because $F_{n-k}(N_+,N_-)$ is homogeneous of degree $n-k$, we have
\begin{equation}
    \fft{d}{dt}F_{n-k}(N_+,N_-)=(n-k)\dot\Phi F_{n-k}(N_+,N_-)+\bar F_{n-k}(N_+,N_-),
\end{equation}
where 
\begin{equation}
\label{F bar}
    \bar{F}_{n-k}(N_+,N_-) = \left.\dot{F}_{n-k}(N_+,N_-)\right|_{\dot{N}_+ \to MN_+, \dot{N}_- \to -MN_-}.
\end{equation}
As a result, (\ref{eq:totder}) becomes
\begin{equation}
    \fft{d}{dt}\left(e^{-\Phi}\dot\Phi^{k-1}F_{n-k}(N_+,N_-)\right)=e^{-\Phi}\left((n-2)\dot\Phi^kF_{n-k}(N_+,N_-)+\dot\Phi^{k-1}\bar F_{n-k}(N_+,N_-)\right),
\end{equation}
\end{widetext}
which allows us to make the integration by parts substitution
\begin{equation}
    \dot\Phi^kF_{n-k}(N_+,N_-)\to-\fft1{n-2}\dot\Phi^{k-1}\bar F_{n-k}(N_+,N_-).
\label{eq:ibpr}
\end{equation}
This can then be used recursively to eliminate all powers of $\dot\Phi$, leaving only canonical terms of the form $F_n(N_+,N_-)$.

We are thus left with a canonical basis consisting of products of traces of $N_+$ and $N_-$ with the exception that $\Tr(L^2)$ has been eliminated.  As we have fully used the complete set of equations of motion, no further simplifications can be obtained through on-shell field redefinitions.  One interesting observation is that, by eliminating all higher derivatives, all $n$-derivative couplings written in this manner are homogeneous of degree $n$ in the $N_+$ and $N_-$ fields.  In particular, while the reduction of linearized $\alpha'^3R^4$ leads to a quartic coupling of the form $(\dot N_\pm)^4$, use of (\ref{eq:Ndotelim}) transforms this into an eight-point coupling of the form $( N_\pm)^8$ which can no longer be seen at the level of the four-point function in the dimensionally reduced theory.

In a T-duality invariant theory, the resulting terms $F_n(N_+,N_-)$ must be expressible in terms of traces of even powers of $\dot{\mathcal S}$ as in (\ref{eq:TrSdn}).  In order for this to happen, the sequence of $N_+$ and $N_-$ in each trace must arrange themselves in alternating order according to (\ref{eq:Tdi}).  This, of course, provides a non-trivial test of T-duality, as noted in \cite{Meissner:1996sa,Hohm:2015doa,Hohm:2019jgu,Codina:2020kvj,Garousi:2021ikb,Garousi:2021ocs}.

\section{Higher curvature invariants and the torsionful connection}
\label{sec:hciR}

So far, we have outlined the general cosmological reduction and described a canonical basis of one-dimensional higher-derivative couplings.  Of course, our aim is to start with a higher-derivative action in ten dimensions and study its reduction.  Focusing only on the NSNS sector, gauge invariant higher-derivative couplings can be expressed in terms of the Riemann tensor $R_{\mu\nu}{}^{\rho\sigma}$, three-form field strength $H_{\mu\nu\rho}$ and the ten-dimensional dilaton $\phi$, along with their covariant derivatives.

From a stringy point of view, it is natural to introduce the torsionful connection $\Omega_\pm=\Omega\pm\fft12H$, given in components in (\ref{eq:Omega+}).  The resulting Riemann tensor computed from $\Omega_\pm$ is then
\begin{equation}
    R_{\mu\nu}{}^{\rho\sigma}(\Omega_\pm)=R_{\mu\nu}{}^{\rho\sigma}(\Omega)\pm\nabla_{[\mu}H_{\nu]}{}^{\rho\sigma}+\fft12H_{[\mu}{}^{\rho\gamma}H_{\nu]\gamma}{}^\sigma,
\label{eq:Rplus}
\end{equation}
which generalizes the linearized expression (\ref{eq:R+lin}).  Tree-level higher curvature invariants can then be written in the form
\begin{equation}
    e^{-2\phi}\mathcal F(R_{\mu\nu}{}^{\rho\sigma}(\Omega_\pm),H_{\mu\nu\rho},\partial_\mu\phi).
\label{eq:calFterm}
\end{equation}
The reduction of couplings not involving additional covariant derivatives is straightforward, and follows from the reduction ansatz (\ref{graviton reduction}).  For the antisymmetric tensor and dilaton, we have
\begin{align}
    H_{\mu\nu\rho}:\qquad\hphantom{\partial_t\phi} H_t{}^i{}_j&=M^i{}_j,\nn\\
    \partial_\mu\phi:\qquad\hphantom{H_t{}^i{}_j} \partial_t\phi&=\ft12\dot\Phi+\ft14\Tr N_{+}=\ft12\dot\Phi+\ft14\Tr N_{-},
\label{eq:Hpd}
\end{align}
while reduction of the torsionful Riemann tensor takes the elegant form
\begin{subequations}
\begin{align}
\label{riemann with torsion}
        \tensor{R}{^{ij}_{kl}}(\Omega_\pm)&=\frac{1}{4}\left(\tensor{N}{^{i}_{k \pm}}\tensor{N}{^{j}_{l \pm}}-\tensor{N}{^{i}_{l \pm}}\tensor{N}{^{j}_{k \pm}} \right),\\
        \tensor{R}{^{ti}_{tj}}(\Omega_\pm) &=
         \frac{1}{4}\left(2\tensor{\dot N}{^{i}_{j \pm}} + \tensor{N}{^{i}_{k}_{\mp}}\tensor{N}{^{k}_{j}_{\pm}}\right).
\end{align}
\end{subequations}
Since we will ultimately make use of on-shell field redefinitions, we allow ourselves to immediately simplify the mixed time-space Riemann expression using (\ref{eq:Ndotelim}) to eliminate $\dot N_\pm$, with the result
\begin{equation}
    \tensor{R}{^{ti}_{tj}}(\Omega_\pm) =
        \frac{1}{4}\left(2\dot\Phi\tensor{N}{^{i}_{j \pm}} + \tensor{N}{^{i}_{k}_{\pm}}\tensor{N}{^{k}_{j}_{\pm}}\right).
\label{eq:Rtitj}
\end{equation}
Note in particular that this pushes the linearized Riemann expression $\bar R^{ti}{}_{tj}(\Omega_\pm)=\fft12\dot N^i{}_{j\pm}$ to non-linear order.

The combination (\ref{eq:Hpd}), (\ref{riemann with torsion}) and (\ref{eq:Rtitj}), which we summarize as
\begin{align}
    H_t{}^i{}_j&=\ft12(N_+^i{}_j-N_-^i{}_j),\nn\\
    \partial_t\phi&=\ft12\dot\Phi+\ft14\Tr N_+,\nn\\
    \tensor{R}{_{kl}^{ij}}(\Omega_-)=\tensor{R}{^{ij}_{kl}}(\Omega_+)&=\frac{1}{4}\left(\tensor{N}{^{i}_{k+}}\tensor{N}{^{j}_{l+}}-\tensor{N}{^{i}_{l+}}\tensor{N}{^{j}_{k+}} \right),\nn\\
    \tensor{R}{_{tj}^{ti}}(\Omega_-) =\tensor{R}{^{ti}_{tj}}(\Omega_+) &=
    \frac{1}{4}\left(2\dot\Phi\tensor{N}{^{i}_{j+}} + (\tensor{N}{_+}\tensor{N}{_+})^{i}{}_{j}\right),
\label{eq:treduct}
\end{align}
provides a consistent assignment of non-linear order of fields with derivative order.  In particular, $n$-derivative terms in ten dimensions will reduce to $n$-point contact interactions between the fields $\dot\Phi$, $N_+$ and $N_-$.  An important implication of this feature is that this cosmological reduction provides a test of T-duality invariance of $n$-derivative terms only at the level of $n$-point functions in the string effective action.  In particular, while eight-derivative couplings in the type II string can first be seen at the level of the four-point function (\textit{i.e.}\ $\alpha'^3R^4$), a complete test of T-duality using this approach can only be done with knowledge of the non-linear terms up to \textit{e.g.}\ $\alpha'^3H^8$ order.  The reason $\alpha'^3R^4$ is sufficient to provide a test of T-duality in \cite{Codina:2020kvj} is because general covariance requires use of the non-linear Riemann tensor, so that $\alpha'^3R^4$ provides full knowledge of eight-graviton scattering at the eight-derivative level.  A complete test involving the $H$-field, as in \cite{Garousi:2021ikb}, requires the full set of couplings up to eight-point contact terms as determined in \cite{Garousi:2020gio,Garousi:2020lof}.

\subsection{The curvature with torsion and hints of generalized geometry}
\label{sec:hgg}

Higher-derivative terms of the form (\ref{eq:calFterm}) can now be reduced according to (\ref{eq:treduct}).  This does not by itself incorporate all gauge invariant couplings, as we could also introduce additional covariant derivatives.  However, as outlined in the previous section, on-shell field redefinitions can be used to bring such terms to canonical form.

By themselves, the individual components in (\ref{eq:calFterm}) are clearly not T-duality invariant.  However, the reduced Riemann with torsion has several noteworthy features.  Recall that one of the motivations for introducing the torsionful connection comes from the closed-string worldsheet.  Along these lines, we can associate the first ($\mu\nu$) and last ($\rho\sigma$) index pairs on $R^{\mu\nu}{}_{\rho\sigma}(\Omega_+)$ with left and right movers on the worldsheet%
\footnote{Here we have chosen a convention where the NSNS polarization tensor is expanded as $\theta_{\mu\nu}=h_{\mu\nu}+B_{\mu\nu}+(\eta_{\mu\nu}-k_\mu\bar k_\nu-\bar k_\mu k_\nu)\phi$, with the first index corresponding to the left side of the string.  This convention singles out the use of the $\Omega_+$ connection in the following expressions.  Equivalent expressions can of be given using $\Omega_-$ along with $N_-$ if desired.}.
So, for example, the contraction
\begin{widetext}
\begin{equation} \label{t8t8 term}
    t_8t_8R^4\equiv t_{\mu_1\mu_2\cdots\mu_8}t^{\nu_1\nu_2\cdots\nu_8}R^{\mu_1\mu_2}{}_{\nu_1\nu_2}(\Omega_+)R^{\mu_3\mu_4}{}_{\nu_3\nu_4}(\Omega_+)R^{\mu_5\mu_6}{}_{\nu_5\nu_6}(\Omega_+)R^{\mu_7\mu_8}{}_{\nu_7\nu_8}(\Omega_+),
\end{equation}
only has $\mu_i$ indices contracted with $\mu_i$ indices and $\nu_i$ indices contracted with $\nu_i$ indices, as can be seen from the definition of the $t_8$ tensor
\begin{align} \label{t8}
\begin{split}
&t_{8}^{\mu_{1} \mu_{2} \mu_{3} \mu_{4} \mu_{5} \mu_{6} \mu_{7} \mu_{8}}\\&=-2\left(\eta^{\mu_{1} \mu_{4}} \eta^{\mu_{2} \mu_{3}} \eta^{\mu_{5} \mu_{8}} \eta^{\mu_{6} \mu_{7}}+\eta^{\mu_{3} \mu_{6}} \eta^{\mu_{4} \mu_{5}} \eta^{\mu_{1} \mu_{8}} \eta^{\mu_{2} \mu_{7}}+\eta^{\mu_{1} \mu_{6}} \eta^{\mu_{2} \mu_{5}} \eta^{\mu_{3} \mu_{8}} \eta^{\mu_{4} \mu_{7}}\right) \\
&\quad+8\left(\eta^{\mu_{1} \mu_{8}} \eta^{\mu_{2} \mu_{3}} \eta^{\mu_{4} \mu_{5}} \eta^{\mu_{6} \mu_{7}}+\eta^{\mu_{1} \mu_{8}} \eta^{\mu_{2} \mu_{5}} \eta^{\mu_{6} \mu_{3}} \eta^{\mu_{4} \mu_{7}}+\eta^{\mu_{1} \mu_{4}} \eta^{\mu_{2} \mu_{5}} \eta^{\mu_{6} \mu_{7}} \eta^{\mu_{8} \mu_{3}}\right) \\
&\quad+\text { anti-symmetrization of each index pair, with total weight one. }
\end{split}
\end{align}
If we only considered the spatial components $R^{ij}{}_{kl}(\Omega_+)$ in (\ref{eq:treduct}), then the row indices of $N_+$ will contract with other row indices, and likewise column indices of $N_+$ will contract with other column indices.  The transpose relation, (\ref{eq:Ntransp}), will then automatically yield traces of alternating $N_+$ and $N_-$, which are T-duality invariant according to (\ref{eq:Tdi}).  The outcome is that contractions of the spatial components $R^{ij}{}_{kl}(\Omega_+)$ where left-moving indices do not talk with right-moving indices are T-duality invariant by construction.

There are several complications that destroy this simple picture of T-duality invariance, however.  The first is that covariant ten-dimensional expressions built out of Riemann reduce not just on the space components but on the mixed time/space components as well.  A quick look at $R^{ti}{}_{tj}(\Omega_+)$ in (\ref{eq:treduct}) indicates that such terms will always break T-duality invariance by themselves because of the presence of the $(N_+)^2$ term.  Therefore any term built out of only the torsional Riemann tensor can never be T-duality invariant without the inclusion of additional $H$-field and/or dilaton dependent terms as well.  Another issue is that matching with string amplitudes does not give just the $t_8t_8R^4$ term but also terms such as
\begin{equation} \label{epsilonepsilon term}
    \epsilon_8\epsilon_8R^4\equiv-\ft12 \epsilon_{\rho\sigma\mu_1\mu_2\cdots\mu_8}\epsilon^{\rho\sigma\nu_1\nu_2\cdots\nu_8}R^{\mu_1\mu_2}{}_{\nu_1\nu_2}(\Omega_+)R^{\mu_3\mu_4}{}_{\nu_3\nu_4}(\Omega_+)R^{\mu_5\mu_6}{}_{\nu_5\nu_6}(\Omega_+)R^{\mu_7\mu_8}{}_{\nu_7\nu_8}(\Omega_+).
\end{equation}
\end{widetext}
The two $\epsilon$ tensors are equivalent to an antisymmetric $\delta$ function which then contracts left and right-moving indices together.  Such contractions, when applied to the spatial components $R^{ij}{}_{kl}(\Omega_+)$, then naturally give rise to terms of the form $(N_+)^k$ that are by their nature non-invariant under T-duality.

Note that the $\epsilon_8\epsilon_8R^4$ contraction also gives rise to Ricci terms which have the reduction
\begin{align}
    R^i{}_j(\Omega_+)&=\ft14(2\dot\Phi+\Tr N_+)N^i{}_{j+},\nn\\
    R^t{}_t(\Omega_+)&=\ft14(2\dot\Phi\Tr N_++\Tr(N_+^2)),\nn\\
    R(\Omega_+)&=\ft14(4\dot\Phi\Tr N_++(\Tr N_+)^2+\Tr(N_+^2)).
\label{eq:Riccis}
\end{align}
None of these expressions can be made T-duality invariant on their own.  In particular, all Ricci components, as well as the Ricci scalar, include the linear trace, $\Tr N_+=\Tr L$, which explicitly breaks T-duality invariance.  The only additional place where $\Tr L$ enters is through the reduction of the ten-dimensional dilaton, as evident in (\ref{eq:treduct}).  This demonstrates that higher derivative Ricci terms must necessarily be paired together with dilaton terms in order to form T-duality invariant couplings.

The torsionful Ricci scalar term in (\ref{eq:Riccis}) can be field redefined away using the two-derivative equations of motion (\ref{EoM at O(1)}).  To see this, we can take the trace of the ten-dimensional Einstein equation, rewrite it in terms of the torsionful $R(\Omega_+)$ and then combine it with the dilaton equation.  The result is
\begin{equation}
    R(\Omega_+)=-4\partial_\mu\phi\partial^\mu\phi+\ft16H_{\mu\nu\rho}H^{\mu\nu\rho},
\label{eq:teom1}
\end{equation}
which reduces to
\begin{equation}
    R(\Omega_+)=4\dot\phi^2+\ft12\Tr M^2.
\end{equation}
Use of the dilaton reduction in (\ref{eq:treduct}) along with the first substitution in (\ref{eq:N2subs}) then demonstrates that this is on-shell equivalent to the torsionful Ricci scalar expression in (\ref{eq:Riccis}).

Similarly, the torsionful Ricci tensor term can be field redefined away through the torsionful Einstein equation
\begin{equation}
    R_{\mu\nu}(\Omega_+)=-2\nabla_\mu\nabla_\nu\phi-\partial^\lambda\phi H_{\lambda\mu\nu},
\label{eq:teom2}
\end{equation}
which gives
\begin{equation}
    R^t{}_t(\Omega_+)=2\ddot\phi,\qquad R^i{}_j(\Omega_+)=\dot\phi N^i{}_{j+},
\end{equation}
when reduced.  These expressions are on-shell equivalent to those in (\ref{eq:Riccis}) as can be verified through use of the one-dimensional equations (\ref{EoM 1D}).

We now reconsider the nature of higher derivative T-duality invariants of the form (\ref{eq:calFterm}).  By use of the torsionful equations of motion, (\ref{eq:teom1}) and (\ref{eq:teom2}), we can write all such invariants without use of the Ricci tensor or Ricci scalar.  In this case, the only source of linear $\Tr L$ terms is the ten-dimensional dilaton.  Since all terms of this form break T-duality invariance, as they cannot be arranged as alternating traces of $N_{+}$ and $N_{-}$, we see that higher derivative dilaton couplings, if any, are highly constrained.

\subsection{Curvature-squared corrections}

Before considering eight-derivative couplings in type II theory, it is instructive to see how the torsionful Riemann tensor can be assembled to form T-duality invariants.  Consider, for example, the reduction of the Riemann-squared combination $R_{\mu\nu\rho\sigma}(\Omega_+)^2$.  Using the reduction (\ref{eq:treduct}), we find
\begin{align}
    R_{\mu\nu\rho\sigma}(\Omega_+)^2\kern-3em&\nn\\
    &=R^{ij}{}_{kl}(\Omega_+)R_{ij}{}^{kl}(\Omega_+)+4R^{ti}{}_{tj}(\Omega_+)R_{ti}{}^{tj}(\Omega_+)\nn\\
    &=\fft18\left(\left(\Tr(N_+N_-)\right)^2-\Tr(N_+N_-N_+N_-)\right)\nn\\
    &\quad+\dot\Phi^2\Tr(N_+N_-)+\dot\Phi\Tr(N_+^2N_-)+\fft14\Tr(N_+^2N_-^2).
\end{align}
The last line in this expression comes from $R_{titj}(\Omega_+)^2$ and contains the traces $\Tr(N_+N_+N_-)$ and $\Tr(N_+^2N_-^2)$ that explicitly break T-duality invariance.  Note that the first substitution in (\ref{eq:N2subs}) allows us to rewrite this expression as
\begin{align}
    R_{\mu\nu\rho\sigma}(\Omega_+)^2&=6\dot\Phi^4-\ft18\Tr(N_+N_-N_+N_-)\nn\\
    &\quad+\dot\Phi\Tr(N_+^2N_-)+\ft14\Tr(N_+^2N_-^2).
\label{eq:R+sq}
\end{align}
If desired, the terms with $\dot\Phi$ can be put into canonical form using the integration by parts relation (\ref{eq:ibpr}).  Nevertheless, since the one-dimensional dilaton $\Phi$ and its time derivatives are invariant under T-duality, it is easy to see what terms are invariant and non-invariant even without eliminating $\dot\Phi$'s.

Of course, the heterotic string as well as the bosonic string has T-duality invariant $\alpha'R^2$ couplings.  Thus the non-invariant terms in (\ref{eq:R+sq}) must combine with other non-invariant terms to form a complete T-duality invariant higher-derivative coupling in the form of alternating traces of $N_{+}$ and $N_{-}$. The mechanism for how this works is somewhat different between the heterotic and bosonic cases, so we will consider them separately.

\subsection{The heterotic string}

The higher-curvature corrections to the bosonic heterotic string action start at the four-derivative level.  Ignoring the heterotic gauge fields, the gravitational sector action up to $\mathcal O(\alpha')$ takes the form \cite{Metsaev:1987zx,Bergshoeff:1988nn,Bergshoeff:1989de,Chemissany:2007he}
\begin{widetext}
\begin{align}
S_{H}=\int d^{10} x \sqrt{-g} e^{-2 \phi}\left(R+4(\partial \phi)^{2}-\frac{1}{12} \widehat{H}^{2}+\fft18\alpha'R_{\mu\nu\rho\sigma}(\Omega_+)R^{\mu\nu\rho\sigma}(\Omega_+)\right),
\label{eq:lagHet}
\end{align}
where the three-form has a non-trivial Bianchi identity
\begin{equation}
    d\widehat H=-\fft14\alpha'\Tr R(\Omega_+)\wedge R(\Omega_+).
\end{equation}
This form of the $\mathcal O(\alpha')$ correction has a natural description in the language of generalized geometry \cite{Coimbra:2011nw}.

As noted in (\ref{eq:R+sq}), the Riemann-squared term in the action is not T-duality invariant by itself.  The way invariance is restored is hiding in the modified Bianchi identity, or more specifically, in the addition of the Lorentz-Chern-Simons term in $\widehat H$
\begin{align}
\widehat{H}=dB-\fft14\alpha'\omega_{3L}(\Omega_+),
\end{align}
where
\begin{equation}
    \omega_{3L}(\Omega_+)=\Tr(\Omega_+\wedge d\Omega_++\fft23\Omega_+\wedge\Omega_+\wedge\Omega_+).
\label{Chern-Simons term}
\end{equation}
Note that the torsionful connection $\Omega_+$ is itself defined in terms of $\widehat H$.  Hence the $\alpha'$ modifications to $\widehat H$ enter nonlinearly, but can be expanded perturbatively.

Since we are only interested in the $\mathcal O(\alpha')$ contributions, we can use the closed $H=dB$ in the definition of $\Omega_+=\Omega+\fft12H$.  In this case, the expansion of $S_H$ to $\mathcal O(\alpha')$ gives the effective four-derivative Lagrangian
\begin{equation}
    e^{-1}\mathcal L_H^{\partial^4}=\fft18\alpha'e^{-2\phi}\left(R_{\mu\nu\rho\sigma}(\Omega_+)R^{\mu\nu\rho\sigma}(\Omega_+)+\fft13H_{\mu\nu\rho}\omega_{3L}^{\mu\nu\rho}(\Omega_+)\right),
\label{eq:het4d}
\end{equation}
Using the cosmological reduction of the torsionful spin connection
\begin{align}
    \Omega_+^{0a}&=\fft12N^j{}_{i-}e_j^adx^i,\nn\\
    \Omega_+^{ab}&=\fft12(e^{ia}\dot e_i^b-e^{ia}\dot e_i^a+e^{ia}e^{jb}\dot b_{ij})dt,
\end{align}
we can obtain the reduction of the Lorentz Chern-Simons term
\begin{equation}
    \omega_{3L}(\Omega_+)=\fft12 g_{ik}\left(-N^k{}_{l+}\dot N^l{}_{j-}+\fft12N^k{}_{l+}N^l{}_{m-}N^m{}_{j-}\right)dt\wedge dx^i\wedge dx^j.
\end{equation}
As a result, we find
\begin{equation}
    H_{\mu\nu\rho}\omega_{3L}^{\mu\nu\rho}=\ft34\Tr\relax(N_+^3N_-)-\ft34\Tr\relax(N_+^2N_-^2).
\end{equation}
Both of these terms break T-duality invariance.  However, when combined with (\ref{eq:R+sq}), we find
\begin{equation}
    e^{-1}\mathcal L_H^{\partial^4}=\fft18\alpha'e^{-\Phi}\left(6\dot\Phi^4-\ft18\Tr(N_+N_-N_+N_-)+\dot\Phi\Tr(N_+^2N_-)+\ft14\Tr\relax(N_+^3N_-)\right).
\end{equation}
\end{widetext}
Since $\mathcal L_{\partial^4}$ is an effective Lagrangian, we can apply the integration by parts substitution, (\ref{eq:ibpr}), to reduce it to canonical form.  Using the relevant four-derivative substitutions
\begin{align}
    \dot\Phi^4&\to0,\nn\\
    \dot\Phi\Tr(N_+^2N_-)&\to\ft14\Tr(N_+N_-N_+N_-)-\ft14\Tr(N_+^3N_-),
\label{eq:pchet}
\end{align}
we finally obtain, for the heterotic four-derivative correction
\begin{equation}
    e^{-1}\mathcal L_H^{\partial^4}=\fft1{64}\alpha'e^{-\Phi}\Tr(N_+N_-N_+N_-)=\fft1{128}\alpha'e^{-\Phi}\Tr\relax(\dot{\mathcal S}^4).
\label{eq:het4T}
\end{equation}
This expression is now manifestly $O(9,9)$ invariant.

\subsection{The bosonic string}

Just like in the heterotic string, the bosonic string also acquires a curvature-squared correction.  However, in this case, $H=dB$ remains closed, but instead there are additional non-linear couplings with $H$ in the effective action \cite{Metsaev:1987zx}
\begin{widetext}
\begin{align}
S_{B} &=\int d^{10} x \sqrt{-g} e^{-2 \phi}\biggl(R+4(\partial\phi)^2-\fft1{12}H^2\nn\\
&\kern9em+\fft14\alpha'\left(R_{\mu \nu \rho \sigma}^{2}-\frac{1}{2}R_{\mu\nu\rho\sigma}H^{\mu\nu\lambda}H^{\rho\sigma}{}_\lambda+\frac{1}{24} H^4-\frac{1}{8}\left(H_{\mu \nu}^{2}\right)^{2}\right)\biggr),
\label{eq:Sbos}
\end{align}
where $H_{\mu \nu}^{2} \equiv \tensor{H}{_{\mu}^{\rho\sigma}} H_{\nu \rho\sigma}$ and $H^{4} \equiv H_{\mu \nu \rho} \tensor{H}{^{\mu}_{\lambda}^{\sigma}} \tensor{H}{^{\nu}_{\sigma}^{\eta}} \tensor{H}{^{\rho}_{\eta}^{\lambda}}$.

As written here, the curvatures are given without torsion.  To highlight the generalized geometry picture and to connect with the torsionful $R^2$ reduction, (\ref{eq:R+sq}), we first rewrite the Riemann terms using (\ref{eq:Rplus}).  Substituting the Riemann-squared expression
\begin{equation}
    R_{\mu \nu \rho \sigma}(\Omega_+)^{2}
    =
    R_{\mu \nu \rho \sigma}^{2}+\nabla_{\mu} \tensor{H}{^{\nu\rho\sigma}} \nabla_{\nu} \tensor{H}{^{\mu}_{\rho\sigma}}-R_{\mu \nu \lambda \sigma} H^{\mu \lambda \rho} \tensor{H}{^{\nu \sigma}_{\rho}}+\frac{1}{8}\left(H_{\mu \nu}^{2}\right)^{2}-\frac{1}{8} H^{4},
\end{equation}
into (\ref{eq:Sbos}), and noting that $R_{[\mu\nu\rho]\sigma}=0$ for torsion-free Riemann, the four-derivative Lagrangian becomes
\begin{equation}
    e^{-1}\mathcal{L}_{B}^{\partial^4} = \fft14\alpha'e^{-2\phi}\left(
    R_{\mu \nu \rho \sigma}(\Omega_+)^{2} - \nabla_{\mu} \tensor{H}{^{\nu \rho\sigma}} \nabla_{\nu} \tensor{H}{^{\mu}_{\rho\sigma}} +\frac{1}{6} H^{4}
   -\frac{1}{4}\left(H_{\mu \nu}^{2}\right)^{2}\right).
\end{equation}
The $(\nabla H)^2$ term can be rewritten on-shell using integration by parts and the equations of motion (\ref{EoM at O(1)}) with the result
\begin{equation}
    \nabla_{\mu} \tensor{H}{^{\nu \rho\sigma}} \nabla_{\nu} \tensor{H}{^{\mu}_{\rho\sigma}}\to\tensor{R}{_{ \mu \nu\rho\sigma}}\tensor{H}{^{\mu\nu\alpha}}\tensor{H}{^{\rho\sigma}_{\alpha}}-\frac{1}{4}(H_{\rho\nu}^2)^2.
\end{equation}
Making this substitution, and then once again rewriting $R_{\mu\nu\rho\sigma}$ using the torsionful connection finally gives%
\footnote{The bosonic string is invariant under world-sheet parity, so the effective Lagrangian is even in $H$.  In particular, the Lagrangian is symmetric under the interchange $\Omega_+\leftrightarrow\Omega_-$, and it is only by convention that we have chosen to write it using $\Omega_+$.}
\begin{equation}
    e^{-1}\mathcal{L}_{B}^{\partial^4} = \fft14\alpha'e^{-2\phi}\left(R_{\mu \nu \rho \sigma}(\Omega_+)^{2} -  R_{\mu\nu\rho\sigma}(\Omega_+)H^{\mu\nu\alpha}\tensor{H}{^{\rho \sigma}_{\alpha}} -\frac{1}{3} H^{4}\right).
\label{eq:bapm}
\end{equation}
This will be the starting point for the cosmological reduction.

In contrast with the heterotic four-derivative Lagrangian (\ref{eq:het4d}), the Lorentz Chern-Simons form does not appear, and it is straightforward to reduce using (\ref{eq:treduct}).  The result is
\begin{align}
    R_{\mu\nu\rho\sigma}(\Omega_+)H^{\mu\nu\alpha}\tensor{H}{^{\rho \sigma}_{\alpha}}&=\ft12\Tr(MN_+MN_-)+2\dot\Phi\Tr(M^2N_+)+\Tr(M^2N_+^2)\nn\\
    &=\ft12\dot\Phi\Tr(N_+^3)-\ft12\dot\Phi\Tr(N_+^2N_-)+\ft14\Tr(N_+^4)-\ft14\Tr(N_+^3N_-),
\end{align}
and
\begin{align}
    H^4&=3\Tr(M^4)\nn\\
    &=\ft38\Tr(N_+^4)-\ft32\Tr(N_+^3N_-)+\ft34\Tr(N_+^2N_-^2)+\ft38\Tr(N_+N_-N_+N_-).
\end{align}
Combining these reductions with (\ref{eq:R+sq}) then gives
\begin{align}
    e^{-1}\mathcal L_B^{\partial^4}&=\fft14\alpha'e^{-\Phi}\Bigl(
    6\dot\Phi^4+\ft32\dot\Phi\Tr(N_+^2N_-)-\ft12\dot\Phi\Tr(N_+^3)
    -\ft14\Tr(N_+N_-N_+N_-)\nn\\
    &\kern6em-\ft38\Tr(N_+^4)+\ft34\Tr(N_+^3N_-)\Bigr).
\end{align}
Finally, the $\dot\Phi$ terms can be canonicalized using (\ref{eq:pchet}), along with
\begin{equation}
    \dot\Phi\Tr(N_+^3)\to-\ft34\Tr(N_+^4)+\ft34\Tr(N_+^3N_-).
\end{equation}
The result is now explicitly T-duality invariant 
\begin{equation}
    e^{-1}\mathcal L_B^{\partial^4}=\fft1{32}\alpha'e^{-\Phi}\Tr(N_+N_-N_+N_-)=\fft1{64}\alpha'e^{-\Phi}\Tr\relax(\dot{\mathcal S}^4),
\end{equation}
and is in fact identical to the heterotic string result in (\ref{eq:het4T}) up to an overall factor of two.

The equivalence of the cosmologically reduced heterotic and bosonic four-derivative couplings is actually not a surprise.  Although the original unreduced actions are distinct, T-duality invariance demands that the reduced four-derivative action takes the form (\ref{eq:L4Tdi}).  Moreover, the $a_1\Tr\relax(\dot{\mathcal S}^2)^2$ term in (\ref{eq:L4Tdi}) can be removed by a field redefinition, so we are left with only a single invariant, namely $\Tr\relax(\dot{\mathcal S}^4)$.  This brings up an important point in that the cosmologically reduced action may not necessarily retain the complete information of the unreduced theory.  Hence, while T-duality invariance of the one-dimensional theory can be used as a consistency check of the higher derivative couplings, it is in itself insufficient to guarantee uniqueness of the unreduced higher derivative action.  This matches the observation of \cite{Marques:2015vua,Baron:2017dvb,Edelstein:2019wzg} that there is a two-parameter family of T-dual invariant $\mathcal O(\alpha')$ actions that interpolate between the heterotic and bosonic cases.

To make the connection to the two-parameter family of \cite{Marques:2015vua,Baron:2017dvb} more direct, note that the combined two and four derivative Lagrangian from (\ref{eq:Sbos}) and (\ref{eq:bapm}) can be written as
\begin{align}
    e^{-1}\mathcal L_B=e^{-2\phi}\Bigl(&R+4(\partial\phi)^2-\fft1{12}H^2+\fft14\alpha'H^{\mu\nu\rho}(R_{\mu\nu}{}^{\alpha\beta}H_\rho{}^{\beta\alpha}+\fft16H_\mu{}^{\alpha\beta}H_\nu{}^{\beta\gamma}H_\rho{}^{\gamma\alpha})\nn\\
    &+\fft14\alpha'R_{\mu\nu\rho\sigma}(\Omega_+)^2\Bigr).
\end{align}
\end{widetext}
This suggests that we could define a shifted $H$ field
\begin{equation}
    \widetilde H_{\mu\nu\rho}\equiv H_{\mu\nu\rho}-\fft32\alpha'(R_{\mu\nu}{}^{\alpha\beta}H_\lambda^{\beta\alpha}+\fft16H_\mu{}^{\alpha\beta}H_\nu{}^{\beta\gamma}H_\rho{}^{\gamma\alpha}).
\end{equation}
Working only to $\mathcal O(\alpha')$, the effective Lagrangian then takes the compact form \cite{Marques:2015vua,Baron:2017dvb}
\begin{equation}
    e^{-1}\mathcal L_B=e^{-2\phi}\left(R+4(\partial\phi)^2-\fft1{12}\widetilde H^2+\fft14\alpha'R_{\mu\nu\rho\sigma}(\Omega_+)^2\right).
\end{equation}
This expression is similar to that of the heterotic string, namely (\ref{eq:lagHet}), although here the Bianchi identity for $\widetilde H$ involves both of the torsionful connections \cite{Marques:2015vua,Baron:2017dvb}
\begin{align}
    \widetilde H&=dB-\fft14\alpha'\left(\omega_{3L}(\Omega_+)-\omega_{3L}(\Omega_-)\right)\nn\\
    d\widetilde H&=-\fft14\alpha'\left(\Tr R(\Omega_+)\wedge R(\Omega_+)-\Tr R(\Omega_-)\wedge R(\Omega_-)\right).
\label{eq:antiave}
\end{align}
This ``anti-averaging" ensures the preservation of world-sheet parity.

What is special about this anti-averaging procedure is that the right-hand side of the Bianchi identity is cohomologically trivial as it is the difference of two representatives of the same characteristic class.  In other words, we can write
\begin{equation}
    \widetilde H=dB-\fft12\alpha'X_3,\qquad d\widetilde H=-\fft12\alpha'X_4,
\end{equation}
where
\begin{equation}
    X_3=\Tr\left(R\wedge\mathcal H+\fft1{12}\mathcal H\wedge\mathcal H\wedge\mathcal H\right),
\end{equation}
and
\begin{equation}
    X_4=\Tr\left(R\wedge D\mathcal H+\fft14\mathcal H\wedge\mathcal H\wedge D\mathcal H\right).
\end{equation}
Here we have used a shorthand notation where $\mathcal H^{\alpha\beta}\equiv H_\mu{}^{\alpha\beta}dx^\mu$ is written as a one-form and $(D\mathcal H)^{\alpha\beta}=\nabla_\mu H_\nu{}^{\alpha\beta}dx^\mu\wedge dx^\nu$.  Since $X_3$ is globally well-defined and gauge-invariant, $X_4=dX_3$ is an exact four-form.  We are thus free to choose either the closed $H$ or the non-closed $\widetilde H$ when considering the bosonic string.  Finally, note that the use of an anti-averaged Lorentz Chern-Simons term, (\ref{eq:antiave}), for the type II string was considered in Appendix~A of \cite{Coimbra:2014qaa} where it was shown that such a generalized connection cannot be consistently defined in the context of generalized geometry.

\section{On the T-duality completion of the type II \texorpdfstring{$\alpha'^3$}{alpha'**3} correction}
\label{sec:apm3}

We now proceed to reexamine the tree-level eight-derivative terms in the type II effective action.  While a complete T-duality invariant was presented in \cite{Garousi:2020gio,Garousi:2020lof}, and shown to reduce to (\ref{eq:ans}) in \cite{Garousi:2021ikb}, our aim is to reformulate the invariant in a more natural stringy framework. One of the complications of working with higher derivative effective actions is that field redefinitions can often be used to transform the action into different, but physically equivalent forms.  For the case at hand, the couplings in \cite{Garousi:2020gio} were obtained in a particular basis given in \cite{Garousi:2020mqn} and then reformulated using field redefinitions into a new basis given in \cite{Garousi:2020lof} that avoids explicit derivatives of the ten-dimensional dilaton.

We approach T-duality invariance of the action by focusing at each order of $H$ separately starting at $\mathcal{O}(H^{0})$. This is because higher derivative terms of order $H^{2n}$ do not affect the counterterms introduced at $\mathcal{O}(H^{2n-2})$. Note, however, that the converse is not true as field redefinitions explicitly raise but do not reduce the order of $H$ as can be seen from \eqref{eq:Ndotelim} and \eqref{F bar}. For this reason, we can study each order of $H$ separately but can only comment on the T-duality invariance up to that order.

Our first goal is to verify T-duality at order $\mathcal{O}(H^0)$ in which a similar approach was taken in \cite{Codina:2020kvj} by focusing on only the $\mathcal{O}(H^{0})$ order in the cosmologoical reduction. We then verify the $H^{2}R^{3}$ couplings found via string amplitudes in \cite{Liu:2019ses}. For this purpose, we can truncate to $\mathcal{O}(H^2)$ and make comments about the T-duality of the string action only up to that order.

As discussed in section~\ref{sec:hgg}, dilaton and Ricci terms must be partnered together in order to maintain T-duality invariance.  It is easy to check that the basis in \cite{Garousi:2020lof} avoids both Ricci and dilaton couplings, so there is no conflict with T-duality.  However, from direct string computations, it is well known that the pure gravity sector gives rise to a correction of the form
\begin{equation}
    e^{-1}\mathcal{L}_{R^4} \sim t_{8} t_{8} R^{4}_{}-\frac{1}{4} \epsilon_{8} \epsilon_{8} R^{4},
\label{eq:strbas}
\end{equation}
where the couplings are defined in \eqref{t8t8 term} and \eqref{epsilonepsilon term} in the limit of vanishing torsion.  Since the $\epsilon_8\epsilon_8R^4$ contains Ricci terms when expanded,  we see that the basis used in \cite{Garousi:2020lof} (which omits Ricci terms)  differs from the one implicit in (\ref{eq:strbas}).  As a result, making a direct comparison between the T-duality invariant of \cite{Garousi:2020lof} and various direct string computations may be somewhat challenging.

In principle, there ought to exist a set of field redefinitions that maps between the basis in \cite{Garousi:2020lof} and the basis extending (\ref{eq:strbas}).  While we have yet to construct this map, it is possible to make a few general observations.  Starting directly from the quartic effective action of the string, (\ref{eq:quartA}), we first promote the torsion-free Riemann tensors in (\ref{eq:strbas}), to the non-linear Riemann with torsion, (\ref{eq:Rplus}).  We then take
\begin{equation}
    e^{-1}\mathcal{L}_{R(\Omega_+)^4} \sim t_{8} t_{8} R(\Omega_+)^{4}-\frac{1}{4} \epsilon_{8} \epsilon_{8} R(\Omega_+)^{4},
\label{eq:RO+ter}
\end{equation}
and make use of the explicit expressions
\begin{widetext}
\begin{align} \label{t8 t8 expanded}
    \begin{split}
    t_{8} t_{8} R(\Omega_+)^{4}&=
    24\tensor{R}{^{\mu_{6} \mu_{5}}_{\nu_{8} \nu_{7}}}(\Omega_{+}) \tensor{R}{^{\mu_{8} \mu_{7}}_{\nu_{6} \nu_{5}}}(\Omega_{+}) \tensor{R}{_{\mu_{5} \mu_{6}}^{\nu_{5} \nu_{6}}}(\Omega_{+}) \tensor{R}{_{\mu_{7} \mu_{8}}^{\nu_{7} \nu_{8}}}(\Omega_{+})
    \\& \quad +
    12\tensor{R}{^{\mu_{6} \mu_{5}}_{\nu_{6} \nu_{5}}} (\Omega_{+})\tensor{R}{^{\mu_{8} \mu_{7}}_{\nu_{8} \nu_{7}}}(\Omega_{+}) \tensor{R}{_{\mu_{5} \mu_{6}}^{\nu_{5} \nu_{6}}}(\Omega_{+}) \tensor{R}{_{\mu_{7} \mu_{8}}^{\nu_{7} \nu_{8}}}(\Omega_{+})
    \\ & \quad + 
     192\tensor{R}{^{\mu_{4} \mu_{5}}_{\nu_{4} \nu_{5}}}(\Omega_{+}) \tensor{R}{_{\mu_{3} \mu_{4}}^{\nu_{3} \nu_{4}}}(\Omega_{+}) \tensor{R}{_{\mu_{5} \mu_{6}}^{\nu_{5} \nu_{6}}}(\Omega_{+}) \tensor{R}{^{\mu_{6} \mu_{3}}_{\nu_{6} \nu_{3}}}(\Omega_{+})
    \\& \quad +
    384 \tensor{R}{^{\mu_{8} \mu_{3}}_{\nu_{8} \nu_{5}}}(\Omega_{+}) \tensor{R}{_{\mu_{3} \mu_{5}}_{ \nu_{6} \nu_{7}}}(\Omega_{+}) \tensor{R}{^{\mu_{5} \mu_{7}}^{\nu_{5} \nu_{6}}}(\Omega_{+}) \tensor{R}{_{\mu_{7} \mu_{8}}^{\nu_{7} \nu_{8}}}(\Omega_{+})
    \\ & \quad
    -48\tensor{R}{^{\mu_{6} \mu_{5}}_{\nu_{8} \nu_{3}}}(\Omega_{+}) \tensor{R}{^{\mu_{8} \mu_{7}}^{ \nu_{3} \nu_{4}}}(\Omega_{+}) \tensor{R}{_{\mu_{5} \mu_{6}}_{\nu_{4} \nu_{7}}}(\Omega_{+}) \tensor{R}{_{\mu_{7} \mu_{8}}^{\nu_{7} \nu_{8}}}(\Omega_{+})
    \\ & \quad
    -32\tensor{R}{^{\mu_{6} \mu_{5}}_{\nu_{8} \nu_{5}}}(\Omega_{+}) \tensor{R}{^{\mu_{8} \mu_{7}}_{ \nu_{6} \nu_{7}}}(\Omega_{+}) \tensor{R}{_{\mu_{5} \mu_{6}^{\nu_{5} \nu_{6}}}}(\Omega_{+}) \tensor{R}{_{\mu_{7} \mu_{8}}^{\nu_{7} \nu_{8}}}(\Omega_{+})
    \\ & \quad 
   -48 \tensor{R}{_{\mu_{8} \mu_{3}}^{\nu_{6} \nu_{5}}}(\Omega_{+}) \tensor{R}{^{ \mu_{3} \mu_{4}}^{\nu_{8} \nu_{7}}}(\Omega_{+}) \tensor{R}{_{\mu_{4} \mu_{7}}_{\nu_{5} \nu_{6}}}(\Omega_{+}) \tensor{R}{^{\mu_{7} \mu_{8}}_{\nu_{7} \nu_{8}}}(\Omega_{+})
    \\ & \quad
    -96 \tensor{R}{_{\mu_{8} \mu_{5}}^{\nu_{6} \nu_{5}}}(\Omega_{+}) \tensor{R}{_{ \mu_{6} \mu_{7}}^{\nu_{8} \nu_{7}}}(\Omega_{+}) \tensor{R}{^{\mu_{5} \mu_{6}}_{\nu_{5} \nu_{6}}}(\Omega_{+}) \tensor{R}{^{\mu_{7} \mu_{8}}_{\nu_{7} \nu_{8}}}(\Omega_{+}),
    \end{split}
\end{align}
and
\begin{equation}
\begin{split}
    \epsilon_8\epsilon_8 R(\Omega_+)^4&=-1536 R^{\mu_1 \mu_2 \mu_3 \mu_4}(\Omega_+) R_{\mu_3}{}^{\mu_5}{}_{\mu_1}{}^{\mu_6}(\Omega_+) R_{\mu_4}{}^{\mu_7}{}_{\mu_5}{}^{\mu_8}(\Omega_+) R_{\mu_6 \mu_8 \mu_2 \mu_7}(\Omega_+)\\
    &\quad- 1536 R^{\mu_1 \mu_2 \mu_3 \mu_4}(\Omega_+) R_{\mu_3 \mu_4}{}^{\mu_5 \mu_6}(\Omega_+) R_{\mu_5}{}^{\mu_7}{}_{\mu_1}{}^{\mu_8}(\Omega_+) R_{\mu_6 \mu_8 \mu_2 \mu_7}(\Omega_+)\\
    &\quad+ 768 R^{\mu_1 \mu_2 \mu_3 \mu_4}(\Omega_+) R_{\mu_3}{}^{\mu_5}{}_{\mu_1}{}^{\mu_6}(\Omega_+) R_{\mu_4}{}^{\mu_7}{}_{\mu_2}{}^{\mu_8}(\Omega_+) R_{\mu_6 \mu_8 \mu_5 \mu_7}(\Omega_+)\\
    &\quad+ 96 R^{\mu_1 \mu_2 \mu_3 \mu_4}(\Omega_+) R_{\mu_3 \mu_4}{}^{\mu_5 \mu_6}(\Omega_+) R_{\mu_5 \mu_6}{}^{\mu_7 \mu_8}(\Omega_+) R_{\mu_7 \mu_8 \mu_1 \mu_2}(\Omega_+)\\
    &\quad- 768 R^{\mu_1 \mu_2 \mu_3 \mu_4}(\Omega_+) R_{\mu_3 \mu_4 \mu_1}{}^{\mu_5}(\Omega_+) R_{\mu_5}{}^{\mu_6 \mu_7 \mu_8}(\Omega_+) R_{\mu_7 \mu_8 \mu_2 \mu_6}(\Omega_+)\\
    &\quad+ 48 R^{\mu_1 \mu_2 \mu_3 \mu_4}(\Omega_+) R_{\mu_3 \mu_4 \mu_1 \mu_2}(\Omega_+) R^{\mu_5 \mu_6 \mu_7 \mu_8}(\Omega_+) R_{\mu_7 \mu_8 \mu_5 \mu_6}(\Omega_+)+\cdots.
\end{split}
\label{eq:e8e8R4x}
\end{equation}
\end{widetext}
Note that the ellipses denote Ricci terms that we do not present explicitly due to their length but are included in the computation.

\subsection{T-duality at \texorpdfstring{$\mathcal O(H^0)$}{O(H**0)}}

Following the procedure highlighted in section~\ref{sec:eomfr}, we cosmologically reduce (\ref{eq:RO+ter}) into a canonical basis given in terms of traces of combinations of $N_+$ and $N_-$. We use the Mathematica package xAct to facilitate manipulations of these lengthy expressions \cite{xAct}.  As already noted, the $t_8t_8R(\Omega_+)^4$ expression does not include any Ricci terms and hence does not give rise to any terms proportional to $\Tr(N_+)$.  Nevertheless, it is not by itself T-duality invariant.  In particular, in the limit of vanishing $H$-field, we find
\begin{equation}
    t_8t_8R^4\ \to\ \fft94\Tr(L^8)-6\Tr(L^3)\Tr(L^5)+\fft{51}{16}\left(\Tr(L^4)\right)^2,
\label{eq:t8t8L8}
\end{equation}
in agreement with the results of \cite{Codina:2020kvj}.  The middle term involving traces of odd powers of $L$ explicitly breaks T-duality invariance.

The $\epsilon_8\epsilon_9R(\Omega_+)^4$ term is composed of invariants involving both the Riemann tensor and the Ricci tensor. With the cosmological reduction, the Ricci tensor as defined in \eqref{eq:Riccis} contains $\Tr(N_+)$ and as a consequence gives terms in the action proportional to the one-dimensional time derivative of the dilaton $\Phi$. Since we follow a procedure to remove powers of $\dot{\Phi}$ and $\Tr(L^2)$, the terms proportional to $\Tr(N_{+})$ gives rise to both terms proportional to $\Tr(N_{+})$ and terms that are affiliated to those that appear in the contractions involving the Riemann tensor. In a sense, the Ricci terms spill over and contribute additional terms that are reminiscent of the Riemann contractions. On the other hand, the converse is not true and the contributions coming from contractions involving the Riemann tensor do not involve terms that are proportional to $\Tr(N_{+})$.  In the limit of vanishing $H$-field, we find

\begin{widetext}
\begin{align}
    \epsilon_8\epsilon_8R^4\ &\to\ 45\Tr(L^8)-24\Tr(L^3)\Tr(L^5)-\fft{45}4\left(\Tr(L^4)\right)^2\nn\\
    &\qquad-30(\Tr L)^2\Tr(L^6)+10(\Tr L)^2\left(\Tr(L^3)\right)^2+24(\Tr L)^3\Tr(L^5)\nn\\
    &\qquad-\fft{45}4(\Tr L)^4\Tr(L^4)+4(\Tr L)^5\Tr(L^3)+\fft1{16}(\Tr L)^8.
\label{eq:e8e8L8}
\end{align}
The first line arises from the Riemann only terms in the expansion (\ref{eq:e8e8R4x}) and agrees with the expression found earlier in \cite{Codina:2020kvj}, while the remain lines have various powers of $\Tr(L)$ and arise from Ricci terms.

Combining (\ref{eq:t8t8L8}) and (\ref{eq:e8e8L8}), we obtain
\begin{equation}
    \begin{split}
    e^{-1}\mathcal{L}_{R^4}\ &\to\ 6 \left(\Tr( L^4)\right)^2-9 \Tr (L^8)-\frac{1}{64}(\Tr L)^8-\Tr(L^3) (\Tr L)^5
    \\
    &+\frac{45}{16} \Tr(L^4) (\Tr L)^4-6 \Tr(L^5) (\Tr L)^3-\frac{5}{2} \left(\left(\Tr( L^3)\right)^2-3 \Tr (L^6)\right) \left(\Tr L\right)^2,
    \end{split}
\end{equation}
\end{widetext}
where the first two terms have a natural T-duality invariant formulation in terms of $\Tr\relax (\dot{\mathcal{S}}^4)$ and $\Tr\relax( \dot{\mathcal{S}}^8)$, as given in (\ref{eq:ans}) and obtained first in \cite{Codina:2020kvj}.  The remaining terms are proportional to $\Tr(L)$ and arise from the Ricci contributions in (\ref{eq:e8e8R4x}).  Since they break T-duality invariance, they must be removed by introducing additional couplings involving the dilaton.  This is a clear demonstration that the full eight-derivative invariant must necessarily include tree-level dilaton couplings, provided we work in the field redefinition frame corresponding to (\ref{eq:RO+ter}).  This issue did not arise in the investigations of \cite{Codina:2020kvj,Garousi:2021ikb} as they worked in a field redefinition frame where the Ricci tensor is absent.

\subsection{T-duality at \texorpdfstring{$\mathcal O(H^2)$}{O(H**2)}}

In addition to dilaton couplings, T-duality necessarily involves the $H$-field.  At $\mathcal O(H^0)$, the purely gravitational couplings take the form of (\ref{eq:strbas}), while at $\mathcal O(H^2)$ two types of couplings are present.  The first is of the form $R^2(\nabla H)^2$ which arises from the expansion of the torsional Riemann tensor in (\ref{eq:RO+ter}), while the second is of the form $H^2R^3$.  The latter terms were obtained in \cite{Liu:2019ses} by matching with the tree-level string five-point amplitude, and take the form
\begin{align}
    \begin{split} \label{eq: LH2R3}
       \mathcal{L}_{H^2 R(\Omega_{+})^3}& \sim -2 t_{8} t_{8} H^{2} R\left(\Omega_{+}\right)^{3}-\frac{1}{6} \epsilon_{9} \epsilon_{9} H^{2} R\left(\Omega_{+}\right)^{3}
       \\&\quad + 8 \cdot 4 ! \sum_{i} d_{i} H^{\mu \nu \lambda} H^{\rho \sigma \zeta} \tilde{Q}_{\mu \nu \lambda \rho \sigma \zeta}^{i} +\cdots,
    \end{split}
\end{align}
to be added to (\ref{eq:RO+ter}).  Here the tensors $Q$ are defined by
\begin{equation} \label{eq: H2R3 Q terms}
    \begin{aligned}
        &\tilde{Q}_{\mu \nu \lambda \alpha \beta \gamma}^{1}=R_{\mu \alpha a}{ }^{b} R_{\nu \beta b}{ }^{c} R_{\lambda c \gamma}{ }^{a}, \\
        &\tilde{Q}_{\mu \nu \lambda \alpha \beta \gamma}^{2}=R_{\mu \nu a}{ }^{b} R_{\alpha \beta b}{ }^{c} R_{\lambda c \gamma}{ }^{a}, \\
        &\tilde{Q}_{\mu \nu \lambda \alpha \beta \gamma}^{3}=R_{\mu \nu a}{ }^{b} R_{\lambda \alpha b}{ }^{c} R_{\beta c \gamma}{ }^{a}, \\
        &\tilde{Q}_{\mu \nu \lambda \alpha \beta \gamma}^{4}=R_{\mu a \alpha}{ }^{b} R_{\nu b \beta}{ }^{c} R_{\lambda c \gamma}{ }^{a},
    \end{aligned}
    \qquad
    \begin{aligned}
        &\tilde{Q}_{\mu \nu \lambda \alpha \beta \gamma}^{5}=R_{\mu a b c} R_{\nu \alpha}{ }^{b c} R_{\lambda \beta \gamma}{ }^{a}, \\
        &\tilde{Q}_{\mu \nu \lambda \alpha \beta \gamma}^{6}=R_{\mu a b c} R_{\alpha \beta}{ }^{b c} R_{\nu \lambda \gamma}{ }^{a}, \\
        &\tilde{Q}_{\mu \nu \lambda \alpha \beta \gamma}^{7}=R_{\mu a b c} R_{\nu}{ }^{a}{ }_{\alpha}{ }^{c} R_{\lambda \beta \gamma}{ }^{b}, \\
        &\tilde{Q}_{\mu \nu \lambda \alpha \beta \gamma}^{8}=R_{\mu \nu \alpha \beta} R_{\lambda a b c} R_{\gamma}{ }^{a b c}.
    \end{aligned}
\end{equation}
It was found in \cite{Liu:2019ses} that the coefficients $d_{i}$ are given by
\begin{align} \label{eq: d coefficients from LM paper}
    \begin{split}
        \left\{d_{i}\right\}=k\left\{1,-\frac{1}{4}, 0, \frac{1}{3}, 1, \frac{1}{4},-2, \frac{1}{8}\right\}.
    \end{split}
\end{align}
with the factor $k=1$.

We can now check if these tree-level $H^2R^3$ couplings are compatible with T-duality at order $\mathcal{O}(H^2)$ following a cosmological reduction and utilizing the $N_{\pm}$ matrices.  Since terms with higher powers of $H$ are undetermined, we introduce a complete basis up to order $H^2R^3$ and leave the rest undetermined.  To be specific, we consider a basis with terms up to two powers of $H$, as explicitly written in the Appendix. The only part of the basis that is fixed are the terms in \eqref{eq: H2R3 Q terms} as demanded by the five-point scattering amplitudes. The complete basis for the Lagrangian up to $H^2R^3$ is then given by \eqref{eq: H2R3 Q terms}, \eqref{eq: H2R3 basis part 1} and \eqref{eq: H2R3 basis part 2}.  Note that, while \eqref{eq: H2R3 Q terms} is written with torsion-free Riemann, one could equally well use the torsionful Riemann there, as it only modifies terms beyond $\mathcal O(H^2)$.

By demanding that the final action is only given by traces of alternating $N_{+}$ and $N_{-}$ as this describes the $O(d,d)$ invariant matrices, the $d_i$ coefficients in \eqref{eq: LH2R3} are found to be
\begin{align}
        d_{3}&= \frac{1}{2}\left(4-d_{1}\right),\quad
        d_{4}= \frac{4}{3},\quad
        d_{6}=  \frac{1}{8}\left(-2d_{1}-4d_{2}-2d_{5}+20\right),\nn\\
        d_{7}&= -8,\quad
        d_{8}= \frac{1}{2}.
\label{eq:Tdualitydi}
\end{align}
As shown, T-duality in the cosmological reduction does not uniquely fix all eight coefficients $d_{i}$ and hence is not sufficient in completely determining the couplings for the five-point function. However, the three coefficients $d_{4},d_{7}$ and $d_{8}$ are uniquely determined and match with \eqref{eq: d coefficients from LM paper} only for $k=4$.  The remaining coefficients also agree with \eqref{eq: d coefficients from LM paper}, provided $k=4$.  This validates the results found in \cite{Liu:2019ses}, however with $k=4$, which corrects a normalization error in that reference%
\footnote{This correction factor $k=4$ was also noted in \cite{Liu:2022bfg}.}.

\subsection{T-duality beyond \texorpdfstring{$\mathcal O(H^2)$}{O(H**2)}}

Ideally, one could extend the T-duality results beyond order $H^2R^3$, potentially going all the way to order $H^8$.  However, terms beyond those that can be probed by the tree-level five-point amplitude have yet to be fully explored from a stringy point of view.  The work of \cite{Garousi:2020mqn} demonstrates that it is possible to form a complete gauge invariant basis of eight-derivative couplings of NSNS fields consisting of 872 terms.  Curiously, imposing T-duality invariance for the circle compactification (\textit{i.e.}~under $g_{\mu9}\leftrightarrow b_{\mu9}$ interchange) is sufficient to fix all 872 terms up to one overall coefficient \cite{Garousi:2020gio}.  However, it turns out that $O(9,9)$ invariance under the cosmological reduction is not nearly as rigid.  In particular, the only quantities that show up in the canonical basis for the one-dimensional theory are combinations of traces of $N_+$ and $N_-$.  

At the eight-derivative level, there are 58 independent terms not containing $\Tr(L)$ originating from invariants involving the Riemann tensor and 51 additional terms involving at least one $\Tr(L)$ factor originating from invariants involving the Ricci tensor and/or the dilaton. In this count, we have also included terms with an odd number of $M$ matrices that do not vanish from the trace properties. There are seven of these types of terms:
\begin{equation}
\begin{aligned}
    &\{\Tr(L^2)\Tr(M^2 L M L^2), \Tr(L) \Tr(M^2 L M L^3),\\
    &\quad\Tr(M^2 L^2 M L^3), \Tr(M^2 L M L^4),\Tr(M^2) \Tr(M^2 L M L^2),\\
    &\quad  \Tr(M^4 L^2 M L),  \Tr(M^3 L M^2 L^2)\}.
\end{aligned}
\end{equation}
As there are only 109 independent terms in the reduced theory, the lift of the one-dimensional T-duality invariant (\ref{eq:ans}) is hardly unique.  Nevertheless, one may expect to find constraints of the form (\ref{eq:Tdualitydi}) that can provide a window on the nature of T-duality invariants at higher order.

\section{Conclusion}
\label{sec:conc}

We have shown that $O(9,9)$ invariance of the cosmologically reduced type~II theory can be repackaged via traces of alternating matrices $N_{+}$ and $N_{-}$. This gives us a particular advantage as it can be manifest which terms appear to be T-duality invariant after the torus reduction.

For example, one important observation is that, while use of a torsionful Riemann tensor (\ref{eq:Rplus}) goes a long way towards preserving T-duality, it cannot be the full story, as the mixed component $R^{ti}{}_{tj}(\Omega_+)$ in the reduction explicitly breaks T-duality invariance.  This can also be seen in the gauge field sector where the lower-dimensional Riemann tensor $R_{\alpha\beta}{}^{\gamma\delta}(\Omega_+)$ for a circle reduction has indefinite $\mathbb Z_2$ parity under the interchange of the momentum and winding $U(1)$ fields \cite{Liu:2013dna}.  The implication is that any higher curvature invariant will necessarily include couplings to the $H$-field beyond those given implicitly in the torsionful connection.

Furthermore, whenever the torsionful Ricci tensor or Ricci scalar is involved, the $O(9,9)$ reduction will give rise to non-invariant terms proportional to $\Tr L$.  As these terms can only be cancelled against dilaton terms, there is a close connection between Ricci and dilaton couplings.  Since Ricci-like terms can be removed or shifted around by field redefinitions, several possibilities can arise.  The first is that neither Ricci nor dilaton terms show up, as in the case of \cite{Garousi:2020lof}.  Another possibility is where Ricci and dilaton terms both show up.  However, it is curious that the original basis of \cite{Garousi:2020mqn,Garousi:2020gio} involves dilaton couplings with no Ricci couplings.  The only way this can be T-duality invariant is if the $\Tr L$ terms arising from the dilaton couplings conspire to cancel among themselves.  In fact, this is what happens, as a field redefinition can be performed to remove the dilaton couplings without introducing any Ricci couplings \cite{Garousi:2020lof}.

Since the effective action (\ref{eq:quartA}) is most directly tied to string four-point and five-point functions containing Ricci terms in the expansion of $\epsilon_8\epsilon_8R(\Omega_+)^4$, its natural completion will include dilaton couplings at tree level.  This has potentially interesting implications for S-duality invariance of the type~IIB string.  The natural framework for discussing $SL(2,\mathbb Z)$ is in the Einstein frame, and with the dilaton combined with the RR axion.  The complexified axi-dilaton transforms with U(1) $R$-charge $\pm2$, so terms with an odd number of $\partial_\mu\phi$ couplings will necessarily break U(1).  Such U(1) violating terms are highly constrained and could provide further input in obtaining the T-duality invariant completion of $\alpha'^3R^4$.

Of course, the full T-duality invariant eight-derivative coupling has already been obtained in \cite{Garousi:2020gio,Garousi:2020lof}, albeit in a different field redefinition frame.  Thus it would be fruitful to explicitly work out the field redefinition required to bring the result of \cite{Garousi:2020gio,Garousi:2020lof} into the form of a completion of (\ref{eq:quartA}).  In principle, this can be done by transforming the torsion-free Riemanns into the torsionful Riemann.  However, a variety of integration by parts will be required to rearrange $\nabla H$-type terms.  One would also have to resolve ambiguities in the map $R_{\mu\nu\rho\sigma}(\Omega)$ to either $R_{\mu\nu\rho\sigma}(\Omega_+)$ or $R_{\mu\nu\rho\sigma}(\Omega_-)$ and to reintroduce Ricci terms using the on-shell equations of motion.

Instead of directly addressing the issue of finding the appropriate field redefinition, it may be more fruitful to consider the nature of the redundancies that arise after compactification on $T^9$.  While the basis of gauge-invariant ten-dimensional couplings is highly redundant, we may wonder if there are any properties regarding the structure of these redundancies and how to organize these bases.  Moreover, the couplings of the form $H^2 R^3$ have been obtained in \cite{Liu:2019ses} and we find that using a basis that incorporates these given $H^2R^3$ terms is consistent in the full eight-derivative action. Using this basis, we are able to study T-duality invariance of the action up to $\mathcal{O}(H^2)$. In principle, we can extend this to higher orders in $H$. However, that would require additional invariants involving $H^{4}, H^{6}$ and $H^{8}$ to ensure T-duality at every order of $H$. A combination of working with the known $H^2R^3$ couplings, including a basis of undetermined terms and matching with \cite{Garousi:2020gio,Garousi:2020lof} could lead to a more complete picture of the hidden structure of the higher-derivative action.

Finally, we promote the use of the cosmological reduction on $T^9$ as a means of studying the generalized geometry of string higher derivative corrections.  While generalized geometry and double field theory do not require compactification, many key features are made explicit in the torus reduction.  As noted in the introduction, by reducing on $T^d$, we make explicit the connection of $O(d,d)$ to both the T-duality group and the generalized structure group of the torus.  This can be seen explicitly in the generalized metric (\ref{eq:ggmet}), which is the natural T-duality covariant combination of scalars from the reduction.  Curiously, there does not appear to be a natural $\mathcal O(\alpha'^3)$ invariant in uncompactified double field theory \cite{Hronek:2020xxi}. However, this obstruction disappears when compactified on a torus.

After compactification to one dimension, we have shown that invariants built out of the generalized metric take the form (\ref{eq:Tdi}), which features a trace of alternating $N_+$ and $N_-$ where $N_\pm=g^{-1}(\dot g\pm\dot b)$.  This strongly hints at a geometrical structure to higher-derivative actions, and it would be interesting to see if a similar structure persists in the scalar sector of general reductions on $T^d$ retaining more non-compact dimensions.  T-duality invariants would presumably still be constructed out of derivatives of the generalized metric (\ref{eq:ggmet}).  However, many more possibilities arise in taking derivatives in the $(10-d)$-dimensional spacetime.  Such connections between T-duality and the generalized geometry of string corrections are worth exploring in more detail, and may ultimately shed light on the hidden symmetries of string theory.

\begin{acknowledgements}
We would like to thank Rodrigo de Le\'on Ard\'on and Ruben Minasian for useful discussions. This work was supported in part by the U.S. Department of Energy under grant DE-SC0007859. M.D. is supported by the NSF Graduate Research Fellowship Program under NSF Grant Number: DGE 1256260. 
\end{acknowledgements}

\appendix*
\begin{widetext}
\section{T-duality basis up to \texorpdfstring{$H^2R^3$}{H**2R**3}} \label{appendix: H2R3 basis} 

As we are interested in verifying T-duality for the five-point contact terms of the form $H^2R^3$, we consider a basis of higher derivative counterterms that are only complete up to two powers of $H$, which include
\begin{equation} \label{eq: H2R3 basis part 1}
    \begin{aligned}
        (\nabla \phi)^8: \quad & y_{1}\left(\nabla_{\alpha} \phi \nabla^{\alpha} \phi\right)^4,
        \\ 
        H^2 (\nabla \phi)^6:  \quad &
        x_{2} H_{\gamma  }{}^{\epsilon  \eta  } H_{\delta  \epsilon  \eta  } (\nabla_{}\phi)_{\alpha  } (\nabla_{\alpha}\phi) (\nabla_{\beta }\phi) (\nabla_{\beta}\phi) (\nabla_{\gamma}\phi) (\nabla_{\delta}\phi),
        \\ \\
        (\nabla \phi)^4 R^2:  \quad & \bar{h}_{8} R(\Omega_{+})_{\beta  }{}^{\delta  }{}_{\gamma  }{}^{\epsilon  } R(\Omega_{+})_{\epsilon  }{}^{\eta  }{}_{\delta  \eta  } (\nabla_{}\phi)_{\alpha  } (\nabla_{\alpha}\phi) (\nabla_{\beta}\phi) (\nabla_{\gamma}\phi),
        \\  \quad &
        h_{11}{} R(\Omega_{+})_{\alpha  }{}^{\epsilon  }{}_{\beta  }{}^{\eta  } R(\Omega_{+})_{\gamma  \eta  \delta  \epsilon  }(\nabla_{\alpha}\phi) (\nabla_{\beta}\phi) (\nabla_{\gamma}\phi) (\nabla_{\delta}\phi),
        \\ \\
        (\nabla \phi)^2 R^3:  \quad & \bar{a}_{82}R(\Omega_{+})_{\alpha  }{}^{\gamma  \delta  \epsilon  }R(\Omega_{+})_{\gamma  }{}^{\eta  }{}_{\beta  \delta  }R(\Omega_{+})_{\epsilon  }{}^{}{}_{\eta} (\nabla^{\alpha}\phi) (\nabla^{\beta}\phi),
        \\  \quad &
        a_{39}{}R(\Omega_{+})_{\gamma  }{}^{\eta  }{}_{\beta  }{}^{\theta  }R(\Omega_{+})^{\gamma  \delta  }{}_{\alpha  }{}^{\epsilon  }R(\Omega_{+})_{\epsilon  \theta  \delta  \eta} (\nabla^{\alpha}\phi) (\nabla^{\beta}\phi),
        \\ \quad &
        a_{56}{} R(\Omega_{+})_{\alpha  }{}^{\gamma  \delta  \epsilon  }R(\Omega_{+})_{\eta  \theta  \gamma  \epsilon  }R(\Omega_{+})^{\eta  \theta  }{}_{\beta  \delta  } (\nabla^{\alpha}\phi) (\nabla^{\beta}\phi).
    \end{aligned}
\end{equation}
Moreover, at the $H^2$ order, we have two types of terms: \eqref{eq: H2R3 Q terms} and those involving Ricci
\begin{equation} \label{eq: H2R3 basis part 2}
    \begin{aligned}
        &H^2 R^3: &&& \\
        &\bar w_{198}{} H_{\eta  }{}^{\iota  \lambda  } H_{\theta  \iota  \lambda  } R(\Omega_{+}) R(\Omega_{+})^{\delta  }{}_{}{}^{\epsilon  } R(\Omega_{+})_{\epsilon  }{}^{\eta  }{}_{\delta  }{}^{\theta  },
        & &
        \bar w_{217}{} H_{\delta  }{}^{\iota  \lambda  } H_{\theta  \iota  \lambda  } R(\Omega_{+})^{\beta  }{}_{}{}^{\gamma  } R(\Omega_{+})_{\beta  }{}^{\delta  \epsilon  \eta  } R(\Omega_{+})_{\epsilon  }{}^{\theta  }{}_{\gamma  \eta  },
        \\ &
        \bar w_{219}{} H_{\delta  }{}^{\iota  \lambda  } H_{\eta  \iota  \lambda  } R(\Omega_{+})^{\beta  }{}_{}{}^{\gamma  } R(\Omega_{+})_{\beta  }{}^{\delta  \epsilon  \eta  } R(\Omega_{+})_{\epsilon  }{}^{}{}_{\gamma},
        & &
        \bar w_{273}{} H_{\eta  }{}^{\iota  \lambda  } H_{\theta  \iota  \lambda  } (R(\Omega_{+}))^2 R(\Omega_{+})^{\epsilon  \eta  }{}_{\epsilon  }{}^{\theta  },
        \\ &
        \bar w_{312}{} H_{\beta  \epsilon  }{}^{\lambda  } H_{\gamma  \iota  \lambda  } R(\Omega_{+})^{\beta  }{}_{}{}^{\gamma  } R(\Omega_{+})^{\delta  \epsilon  \eta  \theta  } R(\Omega_{+})_{\eta  }{}^{\iota  }{}_{\delta  \theta  },
        & &
        \bar w_{347}{} H_{\delta  }{}^{\iota  \lambda  } H_{\theta  \iota  \lambda  } R(\Omega_{+}) R(\Omega_{+})^{\delta  }{}_{}{}^{\epsilon  } R(\Omega_{+})^{\eta  \theta  }{}_{\epsilon  \eta  },
        \\ &
        \bar w_{351}{} H_{\delta  \eta  }{}^{\lambda  } H_{\theta  \iota  \lambda  } R(\Omega_{+})^{\beta  }{}_{}{}^{\gamma  } R(\Omega_{+})_{\gamma  }{}^{\delta  }{}_{\beta  }{}^{\epsilon  } R(\Omega_{+})^{\eta  \theta  }{}_{\epsilon  }{}^{\iota  },
        & &
        \bar w_{363}{} H_{\delta  \theta  }{}^{\lambda  } H_{\epsilon  \iota  \lambda  } R(\Omega_{+})^{\beta  }{}_{}{}^{\gamma  } R(\Omega_{+})_{\gamma  }{}^{\delta  }{}_{\beta  }{}^{\epsilon  } R(\Omega_{+})^{\theta  }{}_{}{}^{\iota  },
        \\ \quad &
        \bar w_{369}{} H_{\delta  \theta  }{}^{\lambda  } H_{\epsilon  \iota  \lambda  } R(\Omega_{+}) R(\Omega_{+})^{\delta  }{}_{}{}^{\epsilon  } R(\Omega_{+})^{\theta  }{}_{}{}^{\iota  },
        & &
        \bar w_{373}{} H_{\beta  \theta  }{}^{\lambda  } H_{\epsilon  \iota  \lambda  } R(\Omega_{+})^{\beta  }{}_{}{}^{\gamma  } R(\Omega_{+})^{\delta  \epsilon  }{}_{\gamma  \delta  } R(\Omega_{+})^{\theta  }{}_{}{}^{\iota  },
        \\ \quad &
        \bar w_{386}{} H_{\delta  \eta  \theta  } H_{\epsilon  \iota  \lambda  } R(\Omega_{+})^{\beta  }{}_{}{}^{\gamma  } R(\Omega_{+})_{\gamma  }{}^{\delta  }{}_{\beta  }{}^{\epsilon  } R(\Omega_{+})^{\eta  \theta  \iota  \lambda  },
        & &
        \bar w_{478}{} H_{\beta  \delta  }{}^{\lambda  } H_{\theta  \iota  \lambda  } R(\Omega_{+})^{\beta  }{}_{}{}^{\gamma  } R(\Omega_{+})_{\gamma  }{}^{\delta  \epsilon  \eta  } R(\Omega_{+})^{\theta  \iota  }{}_{\epsilon  \eta  },
        \\ \quad &
        \bar w_{499}{} H_{\beta  \delta  \theta  } H_{\eta  \iota  \lambda  } R(\Omega_{+})^{\beta  }{}_{}{}^{\gamma  } R(\Omega_{+})_{\gamma  }{}^{\delta  \epsilon  \eta  } R(\Omega_{+})^{\theta  \iota  }{}_{\epsilon  }{}^{\lambda  },
        & &
        \bar w_{503}{} H_{\gamma  \delta  \theta  } H_{\eta  \iota  \lambda  } R(\Omega_{+}) R(\Omega_{+})^{\gamma  \delta  \epsilon  \eta  } R(\Omega_{+})^{\theta  \iota  }{}_{\epsilon  }{}^{\lambda  },
        \\ \quad &
        \bar w_{508}{} H_{\beta  \delta  }{}^{\lambda  } H_{\epsilon  \iota  \lambda  } R(\Omega_{+})^{\beta  }{}_{}{}^{\gamma  } R(\Omega_{+})^{\delta  \epsilon  }{}_{\gamma  }{}^{\eta  } R(\Omega_{+})^{\iota  }{}_{\eta},
        & &
        \bar w_{544}{} H_{\beta  \delta  \iota  } H_{\epsilon  \eta  \lambda  } R(\Omega_{+})^{\beta  }{}_{}{}^{\gamma  } R(\Omega_{+})^{\delta  \epsilon  }{}_{\gamma  }{}^{\eta  } R(\Omega_{+})^{\iota  }{}_{}{}^{\lambda  },
        \\ \quad &
        \bar w_{548}{} H_{\beta  \epsilon  \iota  } H_{\gamma  \eta  \lambda  } R(\Omega_{+})^{\beta  }{}_{}{}^{\gamma  } R(\Omega_{+})^{\delta  \epsilon  }{}_{\delta  }{}^{\eta  } R(\Omega_{+})^{\iota  }{}_{}{}^{\lambda  },
        & &
        \bar w_{551} H_{\beta  \delta  \epsilon  } H_{\gamma  \iota  \lambda  } R(\Omega_{+})^{\beta  }{}_{}{}^{\gamma  } R(\Omega_{+})^{\delta  \epsilon  \eta  \theta  } R(\Omega_{+})^{\iota  \lambda  }{}_{\eta  \theta  }.
    \end{aligned}
\end{equation}
The bars on the coefficients of the counterterms denote the expressions that involve the Ricci scalar while the coefficients not containing the bar denote the expressions involving either Riemann or the dilaton. The subscript is an artifact of the total number of invariants for each type of term as computed in xAct but are otherwise arbitrary.

By demanding T-duality, the coefficients in \eqref{eq: H2R3 basis part 1} and \eqref{eq: H2R3 basis part 2} are given by
\begin{equation} \label{eq: H2R3 basis couplings}
    \begin{aligned}
        a_{39}&= 2560, &
        a_{56}&= -2560, &
        \bar{a}_{82}&= 6144, &
        h_{11}&= 11520, &
        \bar{h}_{8}&= -4096, &
        \\
        \bar{w}_{198}&= 1120, &
        \bar{w}_{217}&= \frac{4864}{3}, &
        \bar{w}_{219}&= -1344, &
        \bar{w}_{273}&= -16, &
        \bar{w}_{312}&= 3584, &
        \\
        \bar{w}_{347}&= -270, &
        \bar{w}_{351}&= -\frac{6272}{3}, &
        \bar{w}_{363}&= \frac{1408}{3}, &
        \bar{w}_{369}&= 180, &
        \bar{w}_{373}&= -2496, &
        \\
        \bar{w}_{386}&= \frac{5120}{3}, &
        \bar{w}_{478}&= -\frac{7744}{3}, &
        \bar{w}_{499}&= -2048, &
        \bar{w}_{503}&= \frac{80192}{27}, &
        \bar{w}_{508}&= -\frac{62336}{9}, &
        \\
        \bar{w}_{544}&= \frac{371584}{27}, &
        \bar{w}_{548}&= \frac{6464}{3}, &
        \bar{w}_{551}&= -\frac{20288}{9}, &
        x_{2}&= -2048, &
        y_{1}&= -1024.
    \end{aligned}
\end{equation}
\end{widetext}


%

\end{document}